\documentclass[fleqn,usenatbib]{mnras}

\usepackage{newtxtext,newtxmath}

\usepackage[T1]{fontenc}

\DeclareRobustCommand{\VAN}[3]{#2}
\let\VANthebibliography\thebibliography
\def\thebibliography{\DeclareRobustCommand{\VAN}[3]{##3}\VANthebibliography}

\usepackage{graphicx}	
\usepackage{amsmath}	
\usepackage[normalem]{ulem}

\newcommand{\comment}[1]{}

\newcommand\hln{\bgroup\markoverwith{\textcolor{orange}{\rule[-.5ex]{2pt}{2.5ex}}}\ULon} 

\usepackage{xspace}

\newcommand{\teff}{\ensuremath{T_{\mathrm{eff}}}\xspace}

\newcommand{\feh}{\rm{[Fe/H]}\xspace}
\newcommand{\afe}{[\alpha/{\rm Fe}]\xspace}

\newcommand{\msun}{{{\rm M}_{\odot}}}

\title[RR Lyrae variables in M3]{Multi-wavelength Photometric Study of RR Lyrae Variables in the Globular Cluster NGC 5272 (Messier 3)
}
\author[N. Kumar et al.]{Nitesh Kumar,$^{1}$\thanks{E-mail:nkumar@physics.du.ac.in}
Anupam Bhardwaj,$^{2}$
Harinder P. Singh,$^{1}$
Marina Rejkuba,$^{3}$
Marcella Marconi$^{4}$ and 
\newauthor
Philippe Prugniel$^{5}$
\\
$^{1}$Department of Physics and Astrophysics, University of Delhi, Delhi 110007, India \\ 
$^{2}$Inter-University Center for Astronomy and Astrophysics (IUCAA), Post Bag 4, Ganeshkhind, Pune 411007, India\\
$^{3}$European Southern Observatory, Karl-Schwarzschild Strasse 2, 85748 Garching, Germany\\
$^{4}$INAF-Osservatorio Astronomico di Capodimonte, Salita Moiariello 16, 80131, Naples, Italy\\
$^{5}$Universit\'e de Lyon, Universit\'e Lyon 1, 69622 Villeurbanne; CRAL, Observatoire de Lyon, CNRS UMR 5574, 69561 Saint-Genis Laval, France
}

\date{Accepted XXX. Received YYY; in original form ZZZ}

\pubyear{2023}

\begin{document}
\label{firstpage}
\pagerange{\pageref{firstpage}--\pageref{lastpage}}
\maketitle

\begin{abstract}
We present a comprehensive photometric study of RR Lyrae stars in the M3 globular cluster, utilising a vast dataset of 3140 optical ($UBVRI$) CCD images spanning 35 years from astronomical data archives. We have successfully identified previously known 238 RR Lyrae stars from the photometric data, comprising 178 RRab, 49 RRc, and 11 RRd stars. Multi-band periodogram was used to significantly improve the long-term periods of $65\%$ of RR Lyrae stars in our sample, thanks to the unprecedentedly long temporal coverage of the observations. The light curve templates were used to obtain accurate and precise mean magnitudes and amplitudes of all RR Lyrae variables. We combined optical ($UBVRI$) and near-infrared (NIR, $JHK_{s}$) photometry of RR Lyrae variables to investigate their location in the colour-magnitude diagrams as well as the pulsation properties such as period distributions, Bailey diagrams and amplitude ratios. The Period-Luminosity relations in $R$ and $I$ bands and Period-Wesenheit relations were derived after excluding outliers identified in CMDs. The Period-Wesenheit relations calibrated via the theoretically predicted relations were used to determine a distance modulus of $\mu = 15.04 \pm 0.04 \,{\rm (stats)} \pm 0.19 \,{\rm {(syst.)}}$ mag (using metal-independent $W_{BV}$ Wesenheit) and $\mu = 15.03 \pm 0.04 \,{\rm (stats)} \pm 0.17 \,{\rm {(syst.)}}$ mag (using metal-dependent $W_{VI}$ Wesenheit). Our distance measurements are in excellent agreement with published distances to M3 in the literature. We also employed an artificial neural network based comparison of theoretical and observed light curves to determine physical parameters (mass, luminosity, and effective temperature) for $79$ non-Blazhko RRab stars that agree with limited literature measurements.
\end{abstract}




\section{Introduction}
RR Lyrae stars are low mass (0.5 $\lesssim$ M/$\msun \lesssim$ 0.8), evolved stars \citep[age $\gtrsim$ 10 Gyr, ][]{savino_a_age_2020} located at the intersection of the horizontal branch and classical \emph{instability strip} in the Hertzsprung-Russell diagram. They are in their central helium-burning phase of the evolutionary stage, similar to the intermediate-mass classical Cepheids (3 $\lesssim$ M/$\msun$ $\lesssim$ 13). Due to their well-defined Period-Luminosity relations (PLRs) at infrared wavelengths, first discussed by \cite{longmore_rr_1986} and later studied by many others including \cite{bono_theoretical_2001, catelan_rr_2004-1, sollima_rr_2006, muraveva_new_2015, bhardwaj_rr_2021}, RR Lyrae variables are useful for deriving distances and potentially calibrating the first step of the cosmic distance ladder \citep{beaton_carnegie-chicago_2016, bhardwaj_high-precision_2020}. Additionally, they are valuable for studying stellar evolution and pulsation \citep{catelan_horizontal_2009}, as well as for tracing old stellar populations in their host galaxies \citep{kunder_impact_2018}. 

NGC 5272 (also known as Messier 3 or M3, with R.A. (J2000) = $13^{\rm h}42^{\rm m}11^{\rm s}$ and Dec (J2000) = $+28^{\circ}22'32''$) is a globular cluster located approximately $11.9$ kpc from the Galactic centre, about $10$ kpc from the Sun, and around $9.7$ kpc above the Galactic plane \citep{harris_catalog_1996}. This cluster has a population of approximately 240 RR Lyrae stars, many of which are fundamental mode RR Lyrae (RRab) stars \citep{clement_variable_2001}. M3 is considered a mono-metallic cluster with a mean metallicity of $\feh \sim -1.5$ dex \citep{harris_new_2010}. The period distribution of the RR Lyrae population in M3 shows a sharp peak at a fundamental pulsation period of $\sim 0.55$ days \citep{jurcsik_photometric_2017}, classifying this cluster as a typical Oosterhoff I (OoI) type cluster \citep{oosterhoff_remarks_1939, fabrizio_use_2019}. 

The large population of RR Lyrae stars and a proximity to M3 have already motivated several long-term studies of optical photometric monitoring \citep{kaluzny_optical_1997, corwin_bv_2001, clementini_distance_2003, hartman_bvi_2005, benko_multicolour_2006, jurcsik_long-term_2012, jurcsik_overtone_2015, jurcsik_photometric_2017}, as well as several spectroscopic investigations \citep{sneden_chemical_2004, cohen_abundances_2005, johnson_235_2005, givens_abundance_2016}. \cite{siegel_swift_2015} investigated the RR Lyrae population of M3 at ultraviolet wavelengths. \cite{longmore_globular_1990} first carried out near-infrared (NIR) studies of RR Lyrae stars in M3 and derived Period-Luminosity relation (PLR) in the $K_{s}$-band using 49 variables from the outer region of the cluster. A more recent investigation by \cite{bhardwaj_near-infrared_2020} presented a time-series analysis that included PLRs derived from $233$ RR Lyrae stars in M3, observed in the $J, H,$ and $K_s$ bands. 

Several theoretical studies have focused on M3 variables, aiming to reproduce their observed pulsation properties, focusing on the period distribution of the RR Lyrae population. These studies were carried out by \cite{marconi_rr_2003, catelan_evolutionary_2004, castellani_rr_2005}, and \cite{fadeyev_period_2019}. The study conducted by \cite{catelan_evolutionary_2004} demonstrated that the predicted period distribution, based on canonical horizontal branch models, does not align with observations. In contrast, \cite{castellani_rr_2005} suggested that a bimodal mass distribution would be necessary to replicate the observed period distribution using canonical models. \cite{denissenkov_constraints_2017} used horizontal branch models to investigate the properties of RR Lyrae and non-variable horizontal branch stars in M3. They found that a distance modulus of $\mu$ = 15.02 mag and a reddening value of $E(B-V)=0.013$ mag provided good agreement with the observed properties of these stars. Utilising the full-phased light curves of RR Lyrae stars, \cite{marconi_modeling_2007} employed non-linear pulsation models to accurately simulate the optical light curves constraining physical parameters of RR Lyrae stars in M3.

While M3 has been extensively studied in the past at optical wavelengths, most studies had limited temporal baseline and wavelength coverage in $BVI$ bands \citep[][and references therein]{corwin_bv_2001, hartman_bvi_2005, cacciari_multicolor_2005, benko_multicolour_2006, jurcsik_long-term_2012, jurcsik_photometric_2017}. 
At longer wavelengths, \citet{bhardwaj_near-infrared_2020} provided homogeneous NIR photometry of RR Lyrae variables in M3. A multi-band photometric study of the large sample of RR Lyrae in M3 using a vast dataset covering long time-baseline and spectral-coverage of observations will be useful to explore and analyse the pulsation properties of RR Lyrae stars in unprecedented detail. Homogeneous optical and NIR light curves will also be useful in modelling multi-band light curves, constraining input parameters to stellar pulsation models. In particular, the optical photometry together with NIR photometric mean magnitudes adopted from \citet{bhardwaj_near-infrared_2020} will be useful in deriving multi-band PLRs and optical-NIR PWRs for RR Lyrae based distance measurements. Moreover, the light curves of a considerable proportion of RR Lyrae stars exhibit amplitude and phase modulations over a timescale significantly longer than their primary pulsation period \citep[e.g., in M3 RR Lyrae in][]{jurcsik_blazhko_2018}. This phenomenon is commonly referred to as the Blazhko effect \citep{blazhko_mitteilung_1907, shapley_changes_1916}. Despite being discovered nearly a century ago, the root cause of this phenomenon is not known, emphasizing the need for further investigation using long-term ground and space-based photometric data \citep{molnar_first_2021}.


The structure of the current paper is as follows. In Section \ref{sec:optical_photometry}, we present the optical photometric observations in various standard filter bands. The identification, period determination, template fitting methodology, and period amplitude diagrams are discussed in Section \ref{sec:RR Lyrae stars}. The colour-magnitude diagrams (CMD) of stars in M3 are presented in Section \ref{sec:CMD_M3}. Empirical Period Luminosity (PL) and Period Wesenheit (PW) relations are derived in Section \ref{sec:rrlyrae_diagnostic}, including a determination of the distance to M3. Fourier parameters of RR Lyrae stars of M3 in different optical bands are obtained and discussed in Section \ref{sec:fourier_analysis}, and the physical parameters of non-Blazhko RRab stars are derived using an Artificial Neural Network (ANN), as described in Section \ref{sec:physical_parameters}. Finally, we summarise our work in Section \ref{sec:summary}.

\section{Optical Photometry}\label{sec:optical_photometry}

We provide accurate, homogeneous, and consistently calibrated multi-band $UBVRI$ photometry for the candidate RR Lyrae stars in M3 based on 3140 optical CCD images obtained from public archives. The observational log and the details of different optical data sets employed in this study are provided in Table \ref{tab:observation_log}. The interested readers are referred to \cite{stetson_homogeneous_2019} for details regarding homogeneous photometry of globular clusters using archival imaging datasets.

All images were preprocessed in a standard way, including bias subtraction and flat-fielding. The photometry and calibration were carried out using the DAOPHOT/ALLFRAME software suite \citep{stetson_daophot_1987, stetson_center_1994}. Following the detection of sources and determination of the PSFs in each image, geometric and photometric information for all detected objects was used to derive a self-consisted set of positions and magnitudes for all stars in every image. This is similar to procedures used in the study of Messier 4, and we refer to \cite{stetson_homogeneous_2014} for more details. We note that the datasets from different telescopes, and instruments on different photometric nights were calibrated with respect to standard stars in many different fields, also including atmospheric extinction corrections. The observations that were taken in non-photometric conditions were calibrated with local secondary standards within the images, which themselves were calibrated with the standard stars. All the possible systematic uncertainties were propagated to the photometric measurements, and therefore, these observations are reliably homogenized within their quoted uncertainties.


The calibrated photometry covers a sky area of approximately 57' × 56', enclosing the cluster centre. A total of 3140 images were successfully photometrically calibrated, including 228 $U$-band, 1108 $B$-band, 1215 $V$-band, 198 $R$-band, and 391 $I$-band images, spanning slightly over 35 years. 
The optical time-series photometry of RR Lyrae stars of M3 is provided in Table \ref{tab:optical_photometry}. We have a maximum of 64/277/299/40/96 observations and a minimum of 2/18/16/4/2 observations in UBVRI bands respectively. The median number of observations in U/B/V/R/I filters are 58/249/264/27/66.

\begin{table*}
    \centering
    \caption{Log of the observations of M3 in optical bands. Here, the `Run' column shows unique run labels, `Dates' indicates the observing dates included, and n$_X$ represents the number of images in filter $X$. These filters consist of the standard Landolt filters (U, B, V, R, I), Sloan filters (u, g, r, i), and Stromgren filters (u, b, y). The 'Multiplex' column denotes the number of CCDs in mosaic cameras.}
    \resizebox{\textwidth}{!}{
\begin{tabular}{rlllllllllr}
\hline
No. & Run & Dates & Telescope & Camera & n$_{\rm U}$, n$_{\rm B}$, n$_{\rm V}$, n$_{\rm R}$, n$_{\rm I}$ & n$_{\rm u}$,n$_{\rm b}$,n$_{\rm y}$ & n$_{\rm u}$,n$_{\rm g}$,n$_{\rm r}$,n$_{\rm i}$ & Multiplex \\
    \hline
    
1 & cf84 & 1984 Jun 22 & Maunakea CFHT 3.6m & RCA1 & -, 1, 1, -, - & -, -, - & -, -, -, - & --- \\
2 & kp36 & 1985 Jun 12 & KPNO 0.9m & RCA & -, 4, 4, -, - & -, -, - & -, -, -, - & --- \\
3 & kp4m & 1985 Jun 16-18 & KPNO 4m & RCA & -, 6, 5, -, - & -, -, - & -, -, -, - & --- \\
4 & cf85 & 1985 Jun 20-24 & Maunakea CFHT 3.6m & RCA1 & -, 10, 10, -, - & -, -, - & -, -, -, - & --- \\
5 & jvw & 1986 Mar 29-Apr 09 & La Palma INT 2.5m & RCA & 4, 6, 8, 9, 6 & -, -, - & -, -, -, - & --- \\
6 & km & 1986 May 08 & La Palma INT 2.5m & RCA & -, 1, 2, -, - & -, -, - & -, -, -, - & --- \\
7 & c91ic41 & 1991 Apr 08 & Maunakea CFHT 3.6m & lick1 & -, 2, 2, -, 2 & -, -, - & -, -, -, - & --- \\
8 & cf91 & 1991 Jul 06-07 & Maunakea CFHT 3.6m & Lick2 & -, 2, 2, -, - & -, -, - & -, -, -, - & --- \\
9 & saic & 1992 May 25 & Maunakea CFHT 3.6m & HRCam/saic1 & -, -, 16, 15, - & -, -, - & -, -, -, - & --- \\
10 & h92ic1 & 1992 May 27 & Maunakea CFHT 3.6m & HRCam & -, 7, -, -, - & -, -, - & -, -, -, - & --- \\
11 & cf92 & 1992 Jun 08-09 & Maunakea CFHT 3.6m & RCA4 & -, 3, -, -, - & -, -, - & -, -, -, - & --- \\
12 & bolte & 1994 Apr 13-16 & KPNO 2.1m & t1ka & -, -, 30, -, 31 & -, -, - & -, -, -, - & --- \\
13 & cf94 & 1994 Jun 06 & Maunakea CFHT 3.6m & Loral3 & -, 4, -, -, 4 & -, -, - & -, -, -, - & --- \\
14 & jka & 1995 Mar 27 & La Palma INT 2.5m & TEK3 & -, 3, 8, -, - & -, -, - & -, -, -, - & --- \\
15 & bond9 & 1997 May 08-10 & KPNO 0.9m & t2ka & -, 4, 4, -, 4 & -, -, - & 4, -, -, - & --- \\
16 & arg & 1997 May 31 & La Palma JKT 1m & TEK2 & -, -, 3, -, 6 & -, -, - & -, -, -, - & --- \\
17 & bond11 & 1998 Mar 21-22 & KPNO 0.9m & t2ka & -, 9, 9, -, 9 & -, -, - & 1, -, -, - & --- \\
18 & int9804 & 1998 Apr 12 & La Palma INT 2.5m & WFC & -, 3, -, 4, - & -, -, - & -, -, -, - & --- \\
19 & dmd & 1998 Jun 24 & La Palma JKT 1m & TEK4 & -, -, 2, -, 2 & -, -, - & -, -, -, - & --- \\
20 & tng2 & 2000 Feb 04 & La Palma TNG 3.6m & OIG & 2, 2, 2, -, - & -, -, - & -, -, -, - & x  2 \\
21 & int0005 & 2000 May 24-28 & La Palma INT 2.5m & WFC & -, 3, 3, -, - & -, -, - & -, -, -, - & x  4 \\
22 & jun00 & 2000 Jun 04 & Maunakea CFHT 3.6m & CFH12K & -, -, 2, -, - & -, -, - & -, -, -, - & x 12 \\
23 & cf0102 & 2001 Feb 16-17 & Maunakea CFHT 3.6m & CFH12K & -, 15, 18, -, 12 & -, -, - & -, -, -, - & x 12 \\
24 & not017 & 2001 Jul 11-15 & La Palma NOT 2.6m & CCD7 & -, 9, 7, -, 10 & -, -, - & -, -, -, - & --- \\
25 & int0202 & 2002 Feb 03-04 & La Palma INT 2.5m & WFC & -, 112, 113, -, - & -, -, - & -, -, -, - & x  4 \\
26 & hannah & 2002 Mar 29 & La Palma JKT 1m & SIT2 & -, 5, 11, 10, - & -, -, - & -, -, -, - & --- \\
27 & int1005 & 2010 May 25-30 & La Palma INT 2.5m & WFC & -, 32, 33, 33, 11 & -, -, - & -, -, -, - & x  4 \\
28 & lee3    & 2011 May 24-27 & KPNO 4m  & Mosaic1.1 & -, -, -, -, - & -, 18, 17 & -, -, -, - & x  8 \\
29 & int1202 & 2012 Feb 22-23 & La Palma INT 2.5m & WFC & 16, -, -, -, - & -, -, - & -, -, -, - & x  4 \\
30 & int1204 & 2012 Apr 23-26 & La Palma INT 2.5m & WFC & 10, 9, 9, -, 6 & -, -, - & -, -, -, - & x  4 \\
31 & int1304 & 2013 Apr 13 & La Palma INT 2.5m & WFC & 10, -, -, -, - & -, -, - & -, -, -, - & x  4 \\
32 & spm1405a & 2014 May 21 & S Pedro Martir 0.8 & E2V-4290 & 1, 2, 2, 2, 2 & -, -, - & -, -, -, - & --- \\
33 & dahl & 2014 Jun 19-25 & Maria Mitchell 0.4 & Roper PVCAM & -, -, 3, 3, - & -, -, - & -, -, -, - & --- \\
34 & int1501 & 2015 Jan 30 & La Palma INT 2.5m & WFC & -, 2, 2, -, - & -, -, - & -, -, -, - & x  4 \\
35 & int1502 & 2015 Feb 25-27 & La Palma INT 2.5m & WFC & -, 9, 9, -, - & -, -, - & -, -, -, - & x  4 \\
36 & int1504 & 2015 Apr 24-27 & La Palma INT 2.5m & WFC & -, 16, 17, -, 8 & -, -, - & -, -, -, - & x  4 \\
37 & int1605 & 2016 May 28-Jun 02 & La Palma INT 2.5m & WFC & 1, 11, 11, -, 10 & -, -, - & -, -, -, - & x  4 \\
38 & int1704 & 2017 Apr 03-05 & La Palma INT 2.5m & WFC & -, 6, 6, -, 6 & -, -, - & -, -, -, - & x  4 \\
39 & iac80a & 2017 Jun 17-19 & Teide IAC80 0.8m & CAMELOT & 2, 2, 2, 2, 2 & -, -, - & -, -, -, - & --- \\
40 & spm1801b & 2018 Jan 28 & S Pedro Martir 0.8 & E2V-4240 & 2, 2, 2, 2, 2 & -, -, - & -, -, -, - & --- \\
41 & int1802 & 2018 Feb 22 & La Palma INT 2.5m & WFC & 3, 3, -, -, - & -, -, - & -, -, -, - & x  4 \\
42 & spm1802 & 2018 Feb 26 & S Pedro Martir 2.1 & E2V-4240 & 3, 3, 2, 3, 3 & -, -, - & -, -, -, - & --- \\
43 & int1805 & 2018 May 20-22 & La Palma INT 2.5m & WFC & 6, 2, 6, 2, - & -, -, - & -, -, -, - & x  4 \\
44 & int1806 & 2018 Jun 12 & La Palma INT 2.5m & WFC & 2, 2, -, 2, - & -, -, - & -, -, -, - & x  4 \\
45 & int1906 & 2019 Jun 26-29 & La Palma INT 2.5m & WFC & 5, 2, -, -, - & -, -, - & -, -, -, - & x  4 \\    
    \hline
\end{tabular}
}
\begin{minipage}{\textwidth}
\vspace{0.1cm}
\small \textbf{Notes}. 1. Observers R. D. McClure and D. A. VandenBerg 2. Observer P. B. Stetson 3. Observer P. B. Stetson 4. Observers J. E. Hesser and D. A. VandenBerg 5. Observer "JVW" 6. Observer "KM" 7. Observers H. Richer and G. Fahlman 8. Observers P. B. Stetson and J. E. Hesser 9. Observers Bolte and Hesser 10. Observers Richer and Buonanno 11. Observers P. B. Stetson and M. Bolte 12. Observer M. Bolte 13. Run ID 94IC15, observers D. A. VandenBerg and P. B. Stetson 14. Observer "JKA" 15. Proposal KP2771, observer H. E. Bond; includes observations in Sloan u filter 16. Observer A. Rosenberg 17. Proposal 98A-0351, observer H. E. Bond 18. Details unavailable 19. Run number 4064, observer "DMD" 20. Observers Ortolani, Bragaglia and "admin" 21. Proposal P14, observer Robert Greimel 22. Run ID 00AC14, observers Stetson and Fahlman 23. Run IDs 01AK1 and 01AF43, observers Park, Sohn and Oh;  R. Ibata and C. Pichon 24. Observer Bruntt 25. Proposal P20, observer G. Gilmore 26. Proposal "ING\_Science," observer Javier Mendez 27. Proposal N5, observer vdWerf 28. Proposal ID 2011A-0197, observer J.-W. Lee
29. Proposal C34, observer Carrera 30. Proposal c70, observer Milone 31. Proposal C143, observer M. Monelli 32. Observers Ricci et al. and Ayala-Michel; data contributed by Raul Michel Murillo 33. Observer E. Dahl 34. Proposals CAT\_S and N22W, observers Oscoz and Vats 35. Proposal C115, observer C. Martinez 36. Proposal N2, observer Henk Hoekstra; proposal C115, observer Clara Martinez 37. Proposal C143, observer Monelli 38. Proposals "(16A)C143" and "CAT\_S," observer M. Monelli 39. Observer M. Monelli 40. Observers Omar and Raul; data contributed by Raul Michel Murillo 41. Proposal C14, observer M. Monelli 42. Observers "ing" and Michel; data contributed by Raul Michel Murillo 43. Proposal C14, observer M. Monelli 44. Proposal C14, observers Monelli and Rusakov 45. Proposal C143, observer Monelli.
\end{minipage}
    \label{tab:observation_log}
\end{table*}

\begin{table*}
    \centering
    \caption{Optical time-series photometry of RR Lyrae stars in M3 cluster. The complete table can be accessed online in machine-readable format. }
\begin{tabular}{lllcllllcc}
\hline
   Id & Run Name & Image Name & Band & Filter Flag & HJD & mag & $\sigma_{\rm mag}$ & $\chi$ & Sharp \\
\hline
    V1 & bond9 & obj1801 & U & b & 2450576.8756 & 15.835 & 0.0446 & 1.55  & 0.056 \\
    V1 & bond9 & obj3001 & U & b & 2450577.8346 & 15.320  & 0.0199 & 1.66  & 0.010 \\
    V1 & bond9 & obj4201 & U & b & 2450578.8378 & 16.346 & 0.0347 & 2.26  & 0.106 \\
    V1 & bond9 & obj4205 & U & b & 2450578.8563 & 15.673 & 0.0256 & 2.23  & 0.067 \\
    V1 & bond11 & obj6801 & U & b & 2450894.9046 & 15.370  & 0.0215 & 1.58  & 0.036 \\
    V1 & int1202 & 1m3U30a\_4 & U & a & 2455979.6201 & 15.689 & 0.0136 & 1.52  & 0.170 \\
    \vdots & \vdots & \vdots & \vdots & \vdots &  \vdots &  \vdots &  \vdots &  \vdots &  \vdots \\
    V1 & c91ic41 & 111709o & B & a & 2448354.9710 & 15.568 & 0.0191 & 4.40 & 0.229 \\
    V1 & c91ic41 & 111710o & B & a & 2448354.9777 & 15.611 & 0.0118 & 5.14  & 0.069 \\
    V1 & bond9 & obj1802 & B & a & 2450576.8794 & 15.628 & 0.0317 & 1.93  & 0.103 \\
    \vdots & \vdots & \vdots &  \vdots &  \vdots &  \vdots &  \vdots &  \vdots & \vdots & \vdots \\
    V1 & c91ic41 & 111716o & V & a & 2448354.9901 & 15.413 & 0.0254 & 5.33  & 0.106 \\
    V1 & c91ic41 & 111718o & V & a & 2448354.9972 & 15.457 & 0.0162 & 7.93  & -0.026 \\
    V1 & bond9 & obj1803 & V & a & 2450576.8815 & 15.390  & 0.0257 & 1.64  & 0.106 \\
    \vdots & \vdots & \vdots &  \vdots &  \vdots &  \vdots &  \vdots &  \vdots & \vdots & \vdots \\
    V1 & c91ic41 & 111693o & I & a & 2448354.9374 & 14.876 & 0.0185 & 4.87  & -0.060 \\
    V1 & c91ic41 & 111694o & I & a & 2448354.9464 & 14.896 & 0.0239 & 11.47  & -0.009 \\
    V1 & bond9 & obj1804 & I & a & 2450576.8836 & 15.024 & 0.0326 & 2.12  & 0.111 \\
    \vdots & \vdots & \vdots &  \vdots &  \vdots &  \vdots &  \vdots &  \vdots & \vdots & \vdots \\
    V1 & int1805 & r1388819\_4 & R & a & 2458261.4427 & 15.527 & 0.0108 & 4.12  & 0.488 \\
    V1 & int1806 & r1392560\_4 & R & a & 2458282.4129 & 15.652 & 0.0095 & 1.55  & 0.556 \\
    V1 & int1806 & r1392561\_4 & R & a & 2458282.4145 & 15.647 & 0.0114 & 3.08  & 0.636 \\
    V3 & bond9 & obj1801 & U & b & 2450576.8756 & 15.102 & 0.0264 & 1.37  & 0.026 \\
    V3 & bond9 & obj3001 & U & b & 2450577.8346 & 16.391 & 0.0352 & 1.49  & 0.081 \\
    V3 & bond9 & obj4201 & U & b & 2450578.8378 & 16.323 & 0.0333 & 2.09  & 0.093 \\
    V3 & bond9 & obj4205 & U & b & 2450578.8563 & 16.364 & 0.0269 & 1.53  & 0.081 \\
    V3 & bond11 & obj6801 & U & b & 2450894.9046 & 16.384 & 0.0366 & 1.42  & 0.014 \\
    V3 & int1202 & 1m3U30a\_4 & U & a & 2455979.6201 & 15.930 & 0.0122 & 1.16  & 0.164 \\
    \vdots & \vdots & \vdots &  \vdots &  \vdots &  \vdots &  \vdots &  \vdots & \vdots & \vdots  \\
\hline  
\end{tabular}
\begin{minipage}{\textwidth}
\vspace{0.1cm}
    {\textbf{Note:} The identifier (Id) corresponds to the nomenclature used in the catalogue by \cite{clement_variable_2001}. HJD denotes the Heliocentric Julian Date. Image quality indices $\chi$ and Sharp pertain to the original image. The filter flag `a' signifies observations conducted with the `Landolt' filter, `b' indicates observations initially in the `Sloan' filter and later transformed to Landolt, while `c' denotes observations in the `Stromgren' filter converted to Landolt.
    We note that these transformed Landolt magnitudes are marked with different symbols in the light curves presented in the manuscript.
    }
\end{minipage}

    \label{tab:optical_photometry}   
\end{table*}

Our sample had several photometric measurements in different standard filters (such as Sloan $u, g, r, i,$ and Stroemgren $b, y$ filters), which were transformed to standard Johnson/Kron-Cousins photometric $UBVRI$ bands defined by the equatorial standards of \cite{landolt_ubvri_1992}. While these measurements were valuable for period determination, we will not incorporate them for determining mean magnitudes and amplitudes. This is because their amplitudes are unreliable due to errors introduced during their photometric transformations. These filters do not precisely align with the bandpasses of standard Landolt filters and, as a result, can bias the mean magnitudes and amplitudes when performing template fitting. 


\section{RR Lyrae stars in Messier 3}\label{sec:RR Lyrae stars}
\subsection{Identification of variable stars}\label{sec:RR Lyrae stars:identification}


The identification of RR Lyrae stars in the M3 globular cluster was accomplished using the reference list compiled by \citet[][hereafter CC01, last updated in March 2019]{clement_variable_2001}. This list provided essential information, including coordinates, $V$-band amplitudes, periods, and classifications for most stars. Out of 178 RRab stars classified in CC01, 176 stars were present in our sample, with data missing for V206 and V4s stars. V217 was identified as an RRab star by \cite{siegel_swift_2015} and V265 was identified as an RRab star by \cite{bhardwaj_near-infrared_2020}, both of which were subsequently adopted in our analysis making the total number of RRab stars to 178.
We extracted 49 RRc stars reported by CC01 however a few of them had to be removed due to data quality concerns (see section \ref{sec:CMD_M3}). Moreover, there are 11 double/multi-mode (RRd) variables in our sample, making the total number of RR Lyrae stars 238. Oosterhoff classification and Blazhko type for these stars were adopted following \cite{jurcsik_overtone_2015, jurcsik_photometric_2017}. Among the analysed stars, 90 were Blazhko variables, with 81 RRab, 5 RRc, and 4 RRd type stars being part of this group. 


\subsection{Period determination and distribution}\label{sec:RR Lyrae stars:Period_Distribution}

To test the accuracy of literature periods, we re-determined periods of RR Lyrae stars based on our long temporal baseline data available in multiple bands.
In our analysis, we utilised the Multi-band Lomb Scargle (MBLS) algorithm \citep{vanderplas_periodograms_2015}, a modified version of the traditional Lomb-Scargle (LS) algorithm \citep{lomb_least-squares_1976, scargle_studies_1982}. The MBLS algorithm is designed to handle photometric time-series data available in multiple bands. 

\begin{figure*}
    \centering
    \includegraphics[width=\textwidth]{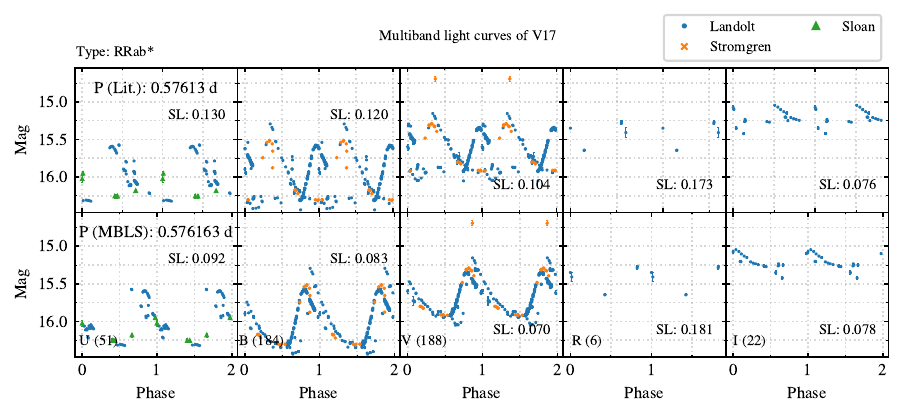}
    \includegraphics[width=\textwidth]{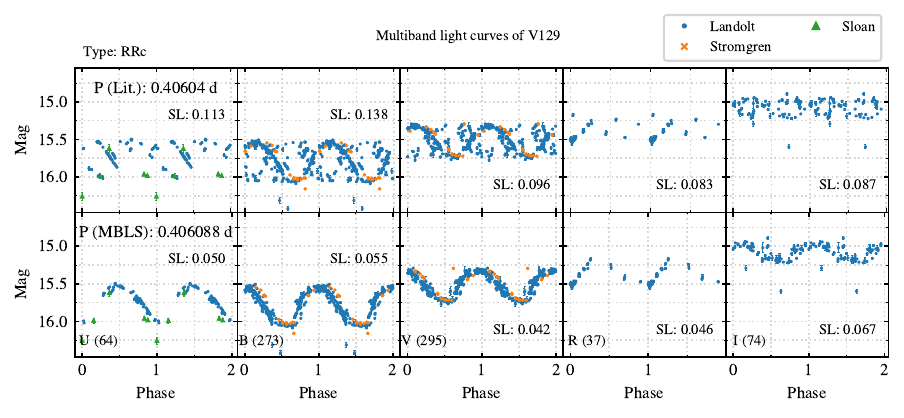}
    \caption{Example phase-folded light curves of V17 and V129 in M3. The upper light curve in each plot is phase folded using the period derived  by \citet{jurcsik_overtone_2015, jurcsik_photometric_2017} and \citet{jurcsik_blazhko-type_2019}. 
    In contrast, the lower light curve in each light curve is phased according to the period obtained through the MBLS algorithm. The period difference between the two methods is 0.000033 days for V17 and 0.000048 for V129, and this discrepancy is reflected in the observed differences in the light curves. Within the plot, SL represents the normalised \emph{string length}, indicating the derived period's accuracy.
    }
%
    
    \label{fig:period_shift_lc}
\end{figure*}

In the classical Lomb-Scargle algorithm, a Fourier transform is performed, generating a power spectrum with candidate frequencies (periodogram). The frequency with the highest power corresponds to the period of the given time-series data. The MBLS algorithm, on the other hand, models the light curves in each band as a Fourier series truncated to arbitrary terms that share the same period and phase among all filter pass bands. This common base model is used across all bands. Subsequently, individual fits are made, and residuals relative to the base model are calculated, resulting in a combined high-power frequency periodogram that considers all bands together, corresponding to the fundamental period. 

\begin{figure}
    \centering
    \includegraphics{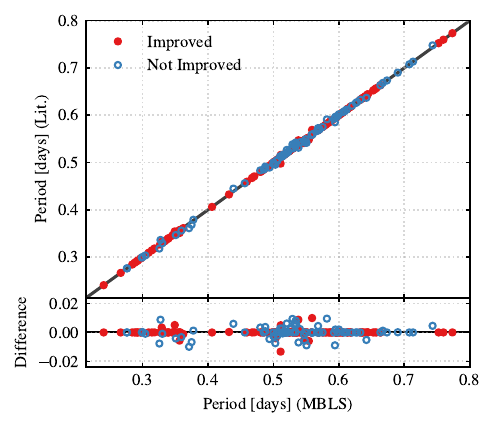}
    \caption{The comparison of Periods derived using Multi Band Lomb Scargle (MBLS) and those reported in literature by \citet{jurcsik_overtone_2015, jurcsik_photometric_2017}, and \citet{jurcsik_blazhko-type_2019}. The periods were considered improved for stars for which the normalised string length is lower using MBLS periods than with the literature periods.}
    \label{fig:period-comparison}
\end{figure}

The Python module {\texttt Gatspy}\footnote{\url{https://github.com/astroML/gatspy}} offers an implementation of the MBLS algorithm for period determination. The typical error bar in the derived periods is $10^{-6}$ days (determined by Monte-Carlo simulation using PERIOD04, \citealt{lenz_period04_2005}). Using the MBLS algorithm, we determined periods for 238 RR Lyrae stars. RR Lyrae stars are known to exhibit both monotonic and random period changes with a typical mean period change rate of the order of 0.01 d/Myr \citep{jurcsik_long-term_2012, szeidl_long-term_2011, li_period_2018}. This suggests that a period change of approximately 3.5 x 10$^{-7}$ days would be expected over a 35-year baseline. The spareness of our light curve sampling results in typical uncertainties of 10$^{-6}$ days in periods and is not ideal for the studies of period changes. To assess the consistency, validity, and accuracy of the derived periods, we performed phase-folding on the light curves using periods obtained through the Multi-Band Lomb Scargle (MBLS) method and those reported by \cite{jurcsik_overtone_2015, jurcsik_photometric_2017, jurcsik_blazhko-type_2019, bhardwaj_near-infrared_2020}, and \cite{clement_variable_2001}. Subsequently, we calculated the \emph{string length} per unit phase. Notably, among the 238 RR Lyrae stars, the MBLS-derived periods exhibited the lowest string length per unit phase for 154 stars. Additionally, periods for 62 RR Lyrae stars were adopted from \cite{jurcsik_overtone_2015, jurcsik_photometric_2017, jurcsik_blazhko-type_2019} as they showed the lower string length per unit phase. Furthermore, periods for 21 stars were adopted from \cite{clement_variable_2001} based on the same criterion, and the period for one RR Lyrae star (V265) was sourced from \cite{bhardwaj_near-infrared_2020}.

A comparison of the phase-folded light curves of stars V17 and V129, with different period is shown in Figure \ref{fig:period_shift_lc}. For V17, the change in period is 0.000033 days when compared with the period derived by \citet{jurcsik_blazhko-type_2019}. It is evident that the observed period derived using MBLS is more accurate for both V17 and V129 as the folded light curve has lower string length, and less scatter.

Figure \ref{fig:period-comparison} shows the comparison of the periods of RR Lyrae stars derived using MBLS and the published periods in the literature \citep{jurcsik_overtone_2015, jurcsik_photometric_2017, jurcsik_blazhko-type_2019}. We note that periods determined in these literature studies are based on dedicated time-series photometry of M3, with well-sampled light curves, but within a much smaller time-baseline than this work. We notice an excellent agreement with the literature periods, where in most cases our derived periods improved the phase light curves based on long-term data. Literature periods were adopted where periods were not determined accurately in our sample. The derived period will be useful for O-C analysis to study the long-term period changes of the RR Lyrae stars in M3 \citep[e.g.,][]{jurcsik_long-term_2012, li_period_2018}.

\begin{figure}
    \centering
    \includegraphics{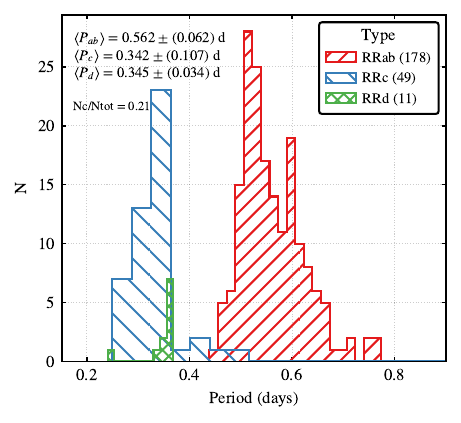}
    \caption{The Period Distribution of RR Lyrae stars in the M3 Globular Cluster.}
    \label{fig:period_distribution}
\end{figure}

The period distribution of RR Lyrae stars is displayed in Fig. \ref{fig:period_distribution}. RRab stars exhibit a wide range of periods, from approximately 0.45 to 0.8 days. However, we do not observe any long-period RR Lyrae stars (P $\geq$ 0.82 days) as they are relatively rare in Galactic globular clusters, with only a few instances found in metal-rich globular clusters such as NGC 6388 and NGC 6441 \citep{pritzl_variable_2001, pritzl_variable_2002, bhardwaj_rr_2022}, as well as in the Galactic field \citep{wallerstein_carbon-rich_2009}.

For typical Oosterhoff I clusters, the ratio between the number of RRc ($\rm N_{\rm c}$) and the total number of RR Lyrae stars ($\rm N_{\rm tot} = \rm N_{\rm ab} + \rm N_{\rm c} + \rm N_{\rm d}$) is approximately 0.29. In contrast, it is around 0.44 for OoII clusters \citep{oosterhoff_remarks_1939, castellani_oosterhoff_1987, caputo_oosterhoff_1990}. In the case of the M3 cluster, the ratio $\rm N_{\rm c} / \rm N_{\rm tot} \sim 0.21$ indicates consistency with an OoI-type cluster. 

M3's RRab stars exhibit a mean period (0.562 days) consistent with Oosterhoff type I (OoI) clusters, while OoII clusters have longer periods, as can be seen in Fig. \ref{fig:period_distribution}. Similarly, the mean period of M3's RRc stars (0.342 days) falls within the range of both OoI and OoII types. The mean metallicity similar to that of M3 ($\textrm{[Fe/H]}\sim -1.5$ dex) typically separates these two Oosterhoff type clusters \citep{catelan_horizontal_2009}.


\subsection{Template fitting}\label{sec:RR Lyrae stars:template_fitting}

The estimation of mean magnitudes for RR Lyrae stars from their sparsely sampled light curves can be achieved by fitting template light curves, requiring fundamental parameters such as period, the epoch of maximum light, and amplitude ratio. Utilising this template fitting approach, it has been possible to achieve a precision of a few hundredths of a magnitude in estimating mean magnitudes, even for light curves with only a few phase points \citep{jones_template_1996, soszynski_mean_2005, inno_new_2015, bhardwaj_near-infrared_2020}.

We used the templates generated by \cite{sesar_light_2009} using Stripe 82 SDSS data to perform a $\chi^2$ minimisation approach for fitting our observed phased light curves. The template collection consists of 11 templates in the $u$-band, 21 templates in the $g$-band, 20 templates in the $r$-band, 20 templates in the $i$-band, and 18 templates in the $z$-band for RRab stars. Similarly, for RRc stars, the template numbers are 1, 2, 2, 2, and 1 in the respective order. In our analysis, we note that $ugriz$ filters are not the same as $UBVRI$, but the approximate light curve shape is not expected to vary significantly at similar wavelengths. Since the template light curves are not available in $UBVRI$, these $ugriz$ templates will best determine mean magnitudes for our RR Lyrae stars. The templates in the `$g$' band were employed for fitting both the $V$ and $B$ band light curves, whereas the templates in the `$u$', `$r$', and `$i$' bands were utilised for fitting the $U$, $R$, and $I$ band light curves, respectively. The templates were sequentially applied to the observed phased light curve corresponding to the appropriate filter. We minimised the $\chi^2$ deviation between the actual magnitude measurements ($m$) and the template fit magnitudes ($m_{\rm fit}$).

\begin{figure*}
    \centering
    \includegraphics{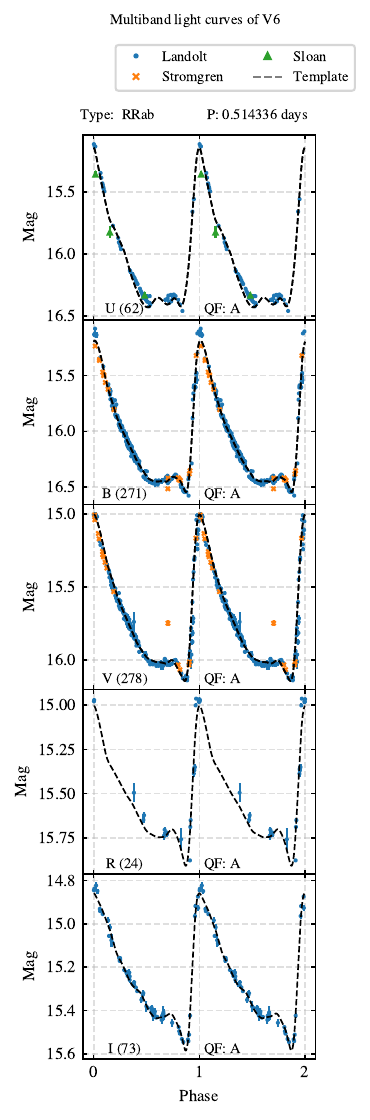}
    \includegraphics{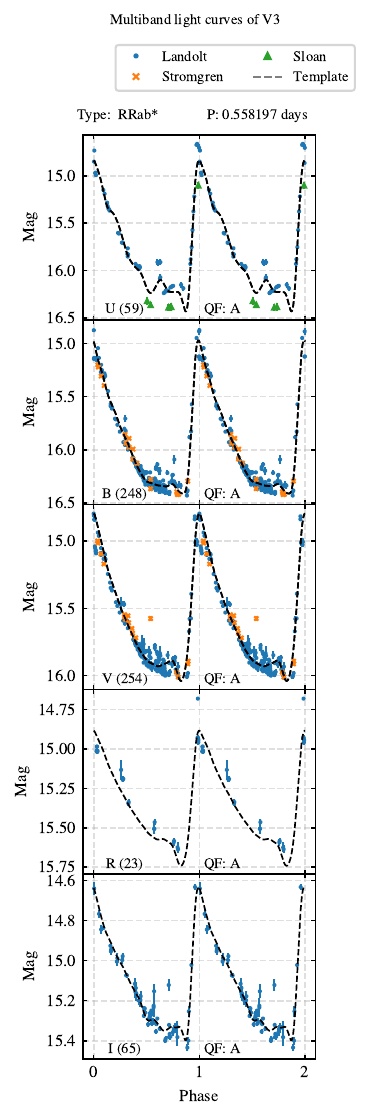}
    \caption{The left panel presents the $U, B, V, R,$ and $I$ band light curves of V6 (non-Blazhko RRab). The period of this star is determined to be 0.514336 days. The folded light curves of the Blazhko RRab star V3 are displayed in the right panel, with a primary period of 0.558197 days. The best-fitted template is represented by the black curve in both plots. The quantity inside the braces adjacent to the filter name in bottom left corner of each plot represents the total number of observations for the corresponding filter band. The corresponding quality flag (QF) is also given at bottom right corner of each light curve.}
    \label{fig:template_fit_rrab}
\end{figure*}

\begin{figure*}
    \centering
    \includegraphics{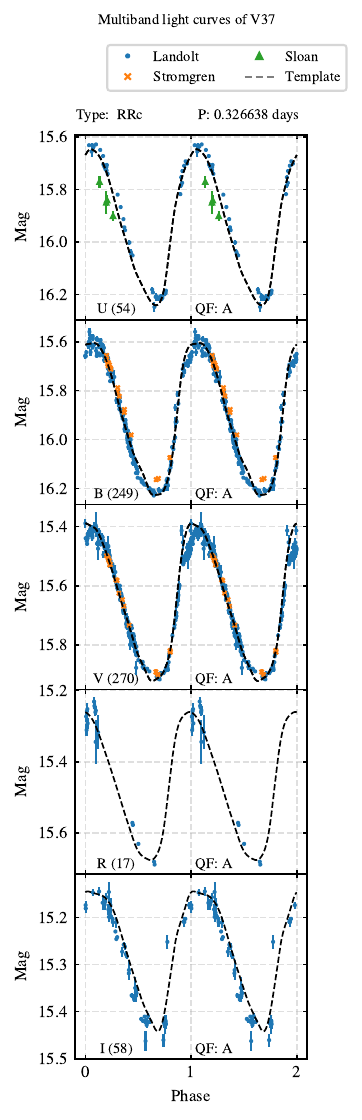}
    \includegraphics{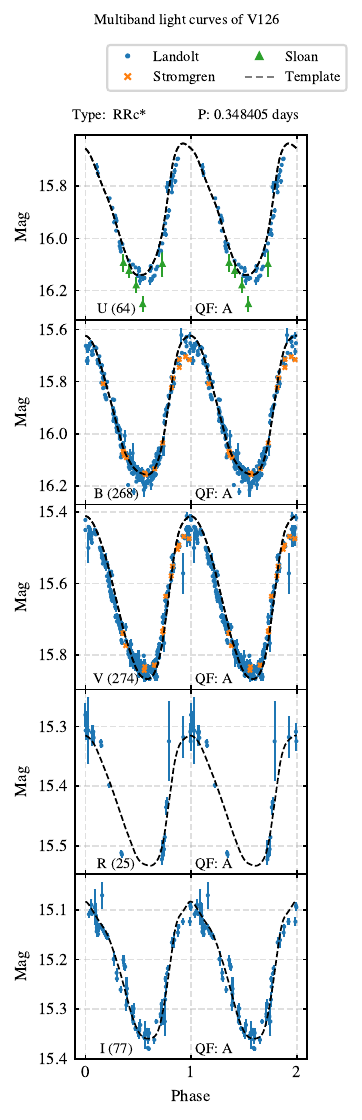}
    \caption{The left panel presents the $U$, $B$, $V$, $R$, and $I$ band light curves of the V37 (non-Blazhko RRc) star. The period of this star is 0.326638 days. The folded light curves of the Blazhko RRc star V126 are displayed in the right panel, with a primary period of 0.348405 days. The adopted nomenclature for this figure is same as Fig. \ref{fig:template_fit_rrab}.}                        
    \label{fig:template_fit_rrc}
\end{figure*}

We phased-folded the observed light curves with the derived period, selecting the zero phase corresponding to the point at which the light curve attains its maximum brightness in the $V$ band. Firstly, we derived a functional from each template light curve using Fourier series, which will be fitted to the observed light curves. In the case of well-sampled $V$ and $B$ band light curves, we solve for the mean-magnitude, amplitude, and a phase offset simultaneously. For the $URI$ band light curves with relatively smaller number of data points, we used amplitude ratios with respect to the $V$ band to scale the amplitudes in these filters \citep{braga_rr_2016}. These scaled amplitudes were allowed to vary by $\pm10\%$ to account for the uncertainties in the amplitude ratios. Most extreme outliers were removed iteratively during the template-fitting process. The best-fitting templates in $UBVRI$ bands were used to determine intensity-averaged mean magnitudes and amplitudes for RR Lyrae variables.

\subsubsection{Light curve quality flags}\label{sec-flags}

\begin{table*}
    \centering
    \caption{Properties of RR Lyrae variables in M3, including their identification (Id), period, intensity averaged mean magnitudes, and amplitudes in the $U, B, V, R,$ and $I$ filters. The Oosterhoff classification, Blazhko variability, and the quality flags (QF) are also provided. The QF values are given in a sequence for the $UBVRI$ filter bands, respectively. The complete table can be accessed online in a machine-readable format.}
    \resizebox{\textwidth}{!}{%
    \begin{tabular}{lcclcllccccccccccc}
\hline
Id & RA$^{\rm c}$ & Dec$^{\rm c}$ & Type & Blazhko$^{\rm b}$ & Period & Oo Type$^{\rm b}$ & <U> & A$_{\rm U}$ & <B> & A$_{\rm B}$ & <V> & A$_{\rm V}$ & <R> & A$_{\rm R}$ & <I> & A$_{\rm I}$ & QF \\
 &  &  &  &  & (days) &  & mag & mag & mag & mag & mag & mag & mag & mag & mag & mag & (UBVRI) \\
\hline
V1 &  13:42:11.12 & +28:20:33.8 & RRab & - & 0.520590$^{\rm b}$ & OoI & 15.78 & 1.31 & 15.90 & 1.38 & 15.62 & 1.18 & 15.41 & 0.81 & 15.17 & 0.68 & AAAAA \\
V3 &  13:42:15.71 & +28:21:41.8 & RRab & Bl & 0.558197$^{\rm a}$ & OoII & 15.70 & 1.59 & 15.85 & 1.44 & 15.53 & 1.25 & 15.34 & 0.86 & 15.09 & 0.77 & AAAAA \\
V5 &  13:42:31.29 & +28:22:20.7 & RRab & Bl & 0.505834$^{\rm a}$ & OoI & 15.97 & 0.94 & 15.90 & 1.42 & 15.58 & 1.26 & 15.55 & 0.56 & 15.16 & 0.83 & AAAAA \\
V6 &  13:42:02.08 & +28:23:41.6 & RRab & - & 0.514336$^{\rm a}$ & OoI & 15.97 & 1.29 & 16.00 & 1.36 & 15.68 & 1.14 & 15.52 & 0.90 & 15.24 & 0.73 & AAAAA \\
V7 &  13:42:11.09 & +28:24:10.2 & RRab & Bl & 0.497423$^{\rm a}$ & OoI & 15.97 & 1.43 & 15.88 & 1.47 & 15.55 & 1.25 & 15.52 & 0.92 & 15.20 & 0.74 & ABBAA \\
\vdots &  \vdots & \vdots & \vdots & \vdots & \vdots & \vdots & \vdots & \vdots & \vdots & \vdots & \vdots & \vdots & \vdots & \vdots & \vdots & \vdots & \vdots \\
V270n &  13:42:11.95 & +28:23:32.7 & RRab & Bl & 0.625850$^{\rm b}$ & - & 15.77 & 0.64 & 15.73 & 0.95 & 15.43 & 0.87 & 15.28 & 0.42 & 14.92 & 0.44 & ABBAA \\
V271 &  13:42:12.18 & +28:23:17.6 & RRab & - & 0.632800$^{\rm c}$ & - & 16.05 & 0.88 & 16.06 & 0.86 & 15.66 & 0.64 & 15.44 & 0.15 & 15.06 & 0.38 & AAAAA \\
V290 &  13:42:21.30 & +28:23:45.0 & RRd & - & 0.240413$^{\rm a}$ & - & 15.91$^\dag$ & 0.25$^\dag$ & 15.85$^\dag$ & 0.54$^\dag$ & 15.67$^\dag$ & 0.07$^\dag$ & 15.53$^\dag$ & 0.22$^\dag$ & 15.38$^\dag$ & 0.30$^\dag$ & ----- \\
V292 &  13:42:11.18 & +28:21:54.0 & RRc & - & 0.296543$^{\rm a}$ & - & 15.81 & 0.36 & 15.77 & 0.37 & 15.61 & 0.29 & 15.45 & 0.21 & 15.32 & 0.21 & AAAAA \\
V299 &  13:41:18.85 & +28:01:57.0 & RRc & - & 0.249200$^{\rm c}$ & - & 15.93$^\dag$ & - & 15.80$^\dag$ & - & 15.60$^\dag$ & - & - & - & 15.42$^\dag$ & - & - \\
\hline
\end{tabular}
    }
    \begin{minipage}{\textwidth}
    \vspace{0.1cm}
    {$^{\rm a}$The adopted period was determined using the Multi-Band Lomb Scargle algorithm. \\
    $^{\rm b}$The adopted period was sourced from \cite{jurcsik_photometric_2017} and \cite{jurcsik_blazhko-type_2019} for RRab stars and from \cite{jurcsik_overtone_2015} for RRc/RRd stars. \\
    $^{\rm c}$The adopted period as well as RA and Dec values were obtained from \cite{clement_variable_2001}. \\
    $^{\rm d}$The adopted period data was extracted from \cite{bhardwaj_near-infrared_2020}. \\
    $^\dag$The intensity-averaged mean magnitudes and amplitudes derived directly from the corresponding light curves.
    } 
\end{minipage}
    \label{tab:period_mag_amp}
\end{table*}

We evaluated the quality of our fitted templates by assigning quality flags based on the mean squared error (MSE) between the best fit template and observed phased light curves. The MSE is calculated as follows:
\begin{equation} \label{eq:rms}
    {\rm MSE} = \frac{\Sigma(m-m_{\rm fit})^2}{N(m)},
\end{equation}
Where m is the light curve magnitudes, $m_{\rm fit}$ is the template magnitudes, and $N(m)$ is the number of measurements of magnitudes in the light curve.

Based on a visual inspection of the template fits and MSE values, we defined three quality flags (QF):
\begin{itemize}
\item \textbf{A}: Phased light curves in a given filter exhibit excellent template fits when MSE is within the range of 0 < MSE $\leq$ 0.02.
\item \textbf{B}: Phased light curves display some scatter, but they still exhibit clear periodicity and reasonably good template fit when MSE falls in the range of 0.02 < MSE $\leq$ 0.04.
\item \textbf{C}: Phased light curves exhibit significant scatter, resulting in poor template fits when MSE is greater than or equal to 0.04.
\end{itemize}

Table \ref{tab:period_mag_amp} shows the derived intensity averaged mean magnitudes and amplitudes in the given filter bands. The Oosterhoff classification, Blazhko variability, and sub-type for each variable were obtained from \citealt{jurcsik_overtone_2015, jurcsik_photometric_2017, jurcsik_blazhko-type_2019} and \citealt{clement_variable_2001}. For cases where we could not find a good template fit, we estimated the mean magnitudes directly from the light curves. The minimum and maximum number of data points utilized to derive the mean magnitudes and amplitudes in $UBVRI$ bands are $17, 72, 75, 4, 12$ and $64, 277, 295, 40, 95$ respectively.

Fig. \ref{fig:template_fit_rrab} shows the light curves of two RRab stars, one being the Blazhko star. The figure also displays the best-fit template. A similar example of light curves and template fits for RRc stars is shown in Fig. \ref{fig:template_fit_rrc}. A few additional light curves and their corresponding template fits are shown in Fig. \ref{fig:apndx_tmplt_fit_rrab} and Fig. \ref{fig:apndx_tmplt_fit_rrc}.

\subsection{Colour-Magnitude Diagrams of M3 with RR Lyrae stars and outlier detection}\label{sec:CMD_M3}
With the intensity averaged mean magnitudes derived using template fitting, we created colour-magnitude diagrams (CMDs) for the M3 stars and the RR Lyrae variables. We have also included infra-red bands ($J$, $H$, and $K_s$) mean magnitudes and amplitudes of M3 RR Lyrae stars from \cite{bhardwaj_near-infrared_2020} for completeness and extending the wavelength range. All the magnitudes were corrected for Galactic extinction $E(B-V) = 0.013$ \citep{vandenberg_constraints_2016} with the conversion factors adopted from \cite{schlegel_maps_1998}. All magnitudes in the subsequent analysis are corrected for reddening. The CMD includes only well-measured stars according to the following photometric quality criteria: sources with $\chi < 1.8$, which quantifies the quality of the fit between the observed star’s profile and the PSF characterising the given image, absolute value of the sharpness parameter ($<0.7$), which selects stars and rejects objects with too sharp (e.g. cosmic ray or spurious detections around saturated sources) or too broad (e.g. blends or extended sources) PSF, as well as sources with magnitude and colour errors $\sigma_V < 0.03$ mag and $\sigma_{B - I} < 0.04$.

\begin{figure*}
    \centering
    \includegraphics{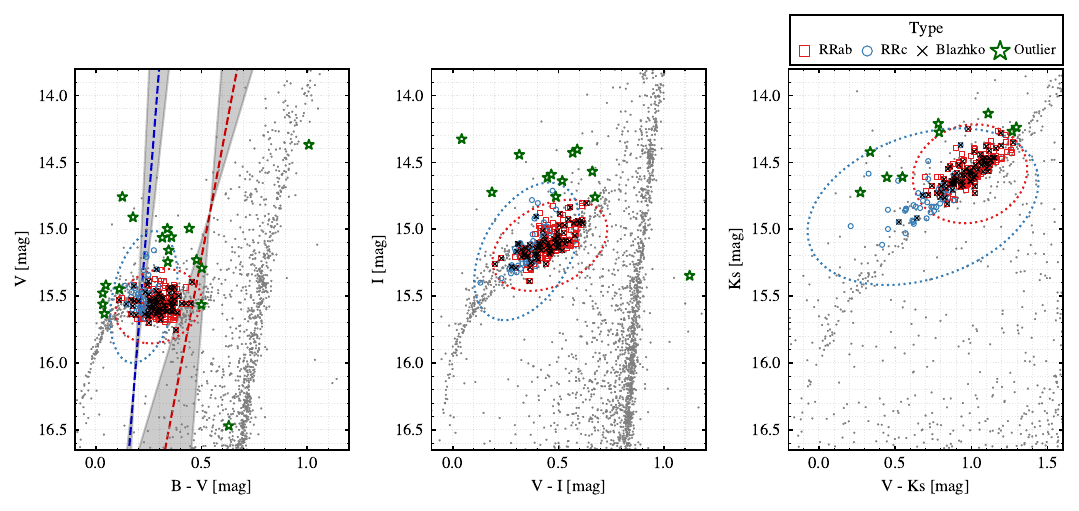}
    \caption{The figure displays the Colour Magnitude diagrams of M3. The grey points represent all the observed stars in the acquired dataset. RRc stars are indicated by blue markers, while RRab stars are represented by red markers. The red and blue dotted lines in the first figure correspond to the theoretical FOBE (first overtone blue edge) and FRE (fundamental mode red edge) obtained using relations from \citet{marconi_new_2015}. The points marked with green star marker are detected outliers in respective CMDs.}
    \label{fig:M3_CMD_1}
\end{figure*}

We have also plotted the theoretical blue (hot) edge for first overtone mode pulsators (FOBE) and the red (cool) edge for fundamental mode pulsators (FRE). These edges were determined using the analytical relations proposed by \cite{marconi_new_2015}, assuming a metallicity of $\feh = -1.50$ (corresponding to Z = 0.00077) with $\afe = 0.20$ and Y = 0.24 \citep{harris_new_2010}. The theoretical instability strip boundaries match the empirical CMD well, but both edges appear redder than the observed distribution of RR Lyrae stars. This slight discrepancy between observed and predicted boundaries can be due to limitations in the current predicted pulsation modelling or the adopted model atmospheres when converting effective temperatures into colours. A better agreement is indeed obtained when the period is adopted in place of the colour. Moreover, the location of the instability edges is known to depend on the efficiency of convection \citep[see e.g.][]{di_criscienzo_rr_2004}. 

To identify potential outliers in these CMDs, we created two ellipses (one for RRab and the other for RRc) in each CMD. The ellipses have the centres at the median of the colour on one axis and the median of the magnitude on the other axis. The eigenvalues and eigenvectors of the covariance matrix of the colour vs magnitude distribution are used to determine the angles of the ellipses. The semi-major and semi-minor axis are defined to be 3 times the standard deviation of the distribution in the respective axis. This ellipse creates an artificial boundary for the distribution of the RRab and RRc stars in the CMDs and any star outside this ellipse is marked as a potential outlier (see Figure \ref{fig:M3_CMD_1}). We created all possible combinations of CMDs using all available bands: $U, B, V, R, I, J, H$, and $K_s$ and then marked potential outliers in each of such plots. We did the statistical analysis of the potential outliers from all CMDs and labelled those stars as true outliers which were outliers in more than 75\% of the CMDs. 3 RRc stars (V259, V297, V298) and 9 RRab stars (V113, V115, V123, V159, V193, V194, V205, V249, V270s) were identified as outlier stars and were removed in the subsequent analysis. Their spurious photometric mean magnitudes and amplitudes are also not provided in Table~\ref{tab:period_mag_amp}. In addition to these, we have excluded stars V4n and V192 from our analysis due to their notably bright V-band magnitudes, as directly calculated from their light curves. These magnitudes are inconsistent with the expected magnitudes of horizontal branch stars within the M3 cluster.

\subsection{Bailey Diagrams and the amplitude ratios}\label{sec:RR Lyrae stars:Bailey_diagram_amplitude_ratio}

\begin{figure}
    \centering
    \includegraphics{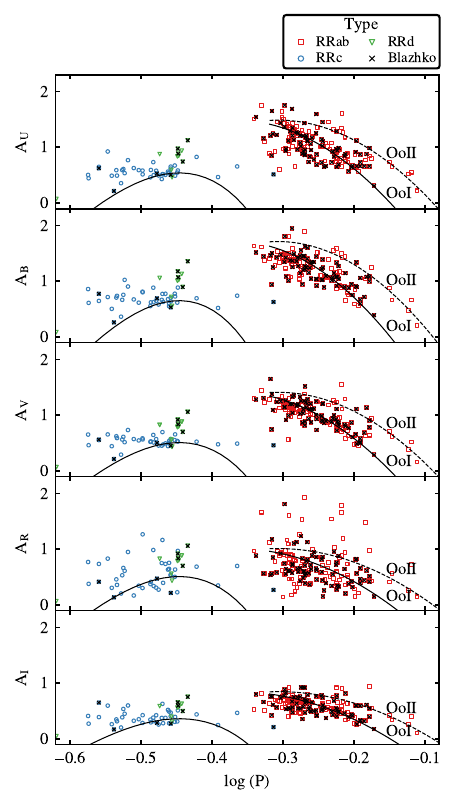}
    \caption{The figure illustrates Bailey's diagram, depicting the period vs amplitude relationship for different filter bands in M3. 
    In addition, the lines are included in the plot for each star type, representing the Oosterhoff-type locus. The equations used to generate these lines are based on \citet{cacciari_multicolor_2005} for RRab OoI and RRab OoII locus was derived using the same equation but with a $\Delta \log {\rm P}$ of 0.06. The relation provided by \citet{kunder_rr_2013} is utilised for RRc stars.}
    \label{fig:P_vs_amplitude}
\end{figure}

The specific pulsation mode of RR Lyrae stars can be determined by their position on a luminosity amplitude-logarithmic period plane, commonly referred to as Bailey's diagram  \citep{bailey_discussion_1902}. Bailey's diagram for the M3 cluster is shown in Fig. \ref{fig:P_vs_amplitude} for all wavelength bands. 
We have plotted a solid line for OoI and a dotted line for OoII RRab stars to represent the locus of Oosterhoff-type stars. These were generated using the following equation provided by \cite{cacciari_multicolor_2005}. For RRab OoI type stars, 
\begin{equation}
    A_{B} = -3.123 - 26.331 (\log P) - 35.853 (\log P)^2,
\end{equation}
and for RRab OoII, the same equation holds with $\Delta \log P = 0.06$. For RRc, we used the relation,
\begin{equation}
    A_V = -9.75 + 57.3 \times P - 80.0 \times P^2,
\end{equation}
given by \cite{kunder_rr_2013}. Additionally, the correlation between the amplitudes of RR Lyrae stars at different wavelengths is evident with a decrease in amplitudes at longer wavelengths. The graph illustrates the well-known property of pulsating stars' amplitudes decreasing as the wavelength increases \citep[see also][]{das_variation_2018, bhardwaj_comparative_2017}.

\begin{figure}
    \centering
    \includegraphics{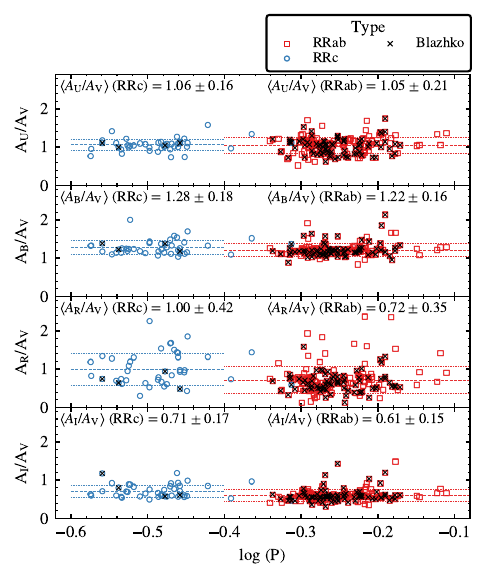}
    \caption{The figure illustrates the distribution of the luminosity amplitude ratio, relative to the $V$ band luminosity amplitude, as a function of the period of the stars.}
    \label{fig:ampl_ratio}
\end{figure}

In line with the methodology employed by \cite{kunder_rr_2013}, \cite{stetson_optical_2014}, and \cite{braga_rr_2016}, we also calculated the amplitude ratios in different bands. Fig. \ref{fig:ampl_ratio} illustrates the obtained amplitude ratios relative to the $V$ band for RRab and RRc stars in M3. We found that the amplitude ratios exhibited similar values (within the error bars) for both types of stars. Moreover, we observed no difference in the amplitude ratio between short-period (P$_{\rm ab} \leq 0.6$ days) and long-period (P$_{\rm ab} \geq 0.6$ days) RRab stars in contrast with findings from NIR light curves \citep{bhardwaj_near-infrared_2020}. The overall amplitude ratio values are in line with literature values from \cite{braga_rr_2016} for $\omega$ Cen and other GCs \citep{di_criscienzo_new_2011, kunder_rr_2013, stetson_optical_2014}. Table \ref{tab:AlAl} provides the estimated amplitude ratios in different bands.

\begin{table}
    \centering
    \caption{Mean amplitude parameter ratio (with respect to $V$ band amplitude) and associated standard deviation for the RRL stars in the M3 globular cluster. }
    \label{tab:AlAl}
    \begin{tabular}{ccccccccc}
    \hline
    {} & \multicolumn{2}{c}{RRab} & \multicolumn{2}{c}{RRc} \\
    {} &   mean &    std &   mean &    std \\
    \hline

$\frac{{\rm A}_{\rm U}}{{\rm A}_{\rm V}}$ & 1.05 & 0.21 & 1.06 & 0.16 \\
$\frac{{\rm A}_{\rm B}}{{\rm A}_{\rm V}}$ & 1.22 & 0.16 & 1.28 & 0.18 \\
$\frac{{\rm A}_{\rm R}}{{\rm A}_{\rm V}}$ & 0.72 & 0.35 & 1.00 & 0.42 \\
$\frac{{\rm A}_{\rm I}}{{\rm A}_{\rm V}}$ & 0.61 & 0.15 & 0.71 & 0.17 \\

    \hline
    \end{tabular}
\end{table}

While \cite{inno_new_2015} proposed a possible dependence of amplitude ratios on metallicity for Cepheids, recent studies conducted on several clusters with diverse metallicities have revealed no clear correlation of amplitude ratios with metallicity \citep{braga_rr_2016} for RR Lyrae stars. Notably, despite their different metallicities, the amplitude ratios for M3 and $\omega$ Cen \citep[derived by][]{braga_rr_2016} are found to be consistent, further supporting the absence of dependence on metallicity.

\section{M3 RR Lyrae: Period-Luminosity, Wesenheit Relations, and A Distance determination}\label{sec:rrlyrae_diagnostic}

\subsection{Period-Luminosity (PL) Relations}\label{sec:PL_relations}
RR Lyrae exhibit a well-defined Period-Luminosity relationship (PLR) in near-infrared wavelength bands \citep[][and references therein]{longmore_rr_1986, bhardwaj_near-infrared_2020} but do not obey any defined relation in optical wavelength bands (UBV) either in empirical data \citep{benkho_blazhko_2011, braga_distance_2015} or in theoretical models \citep{bono_theoretical_2001, catelan_rr_2004, marconi_new_2015}.


\begin{table*}
    \centering
    \caption{Period-Luminosity Relationship for $I$ and $R$ band: ${\rm m_{\lambda}} = {\rm b_{\lambda} \log(P) +  a_{\lambda}}$. }
    \resizebox{\textwidth}{!}{%
\begin{tabular}{ccccccccccccc}
\hline
 & \multicolumn{4}{c}{RRab} & \multicolumn{4}{c}{RRc} & \multicolumn{4}{c}{Global} \\
$\lambda$  & N & b$_\lambda$ & a$_\lambda$ & $\sigma$  & N & b$_\lambda$ & a$_\lambda$ & $\sigma$ &  N & b$_\lambda$ & a$_\lambda$ & $\sigma$ \\
\hline
I & 157 & -1.804$\pm$0.004 & 14.659$\pm$0.001 & 0.107 & 42 & -2.020$\pm$0.007 & 14.162$\pm$0.003 & 0.149 & 192 & -1.239$\pm$0.003 & 14.788$\pm$0.001 & 0.124 \\
R & 159 & -1.453$\pm$0.005 & 14.986$\pm$0.001 & 0.153 & 42 & -1.698$\pm$0.009 & 14.499$\pm$0.004 & 0.166 & 202 & -0.677$\pm$0.003 & 15.158$\pm$0.001 & 0.164 \\

\hline
\end{tabular}
}
    \label{tab:PLR}
\end{table*}

\begin{figure*}
    \centering
    \includegraphics{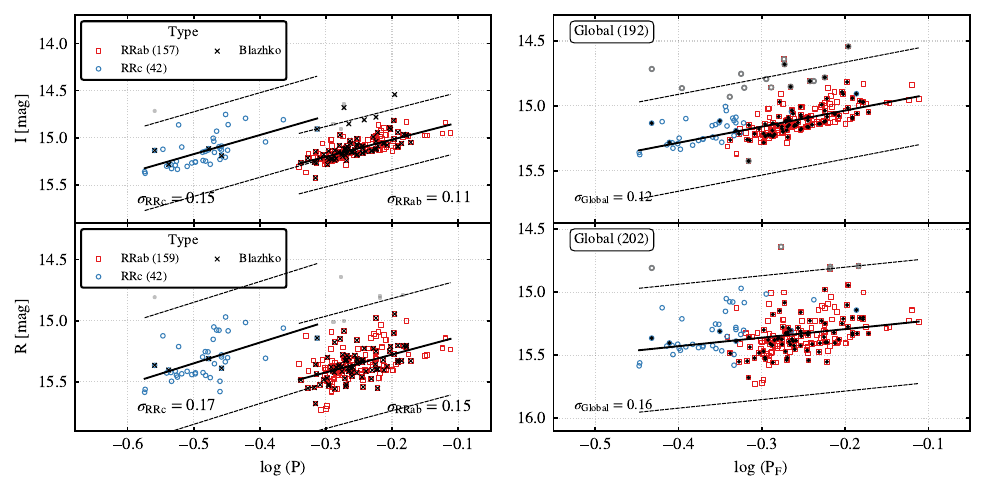}
    \caption{This figure shows the Period Luminosity Relations (PLRs) for the $I$ and $R$ bands. In the right panel, $\rm P_{\rm F}$ represents the period resulting from the conversion of `FO' mode periods into `FU' mode using Equation \ref{eq:fundamentalisation}. The scatter in the PLRs is represented by $\sigma$. The solid line represents the fitted PLR, and the region of $\pm 3 \sigma$ is represented by the dashed lines.}
    \label{fig:PLR_PIR}
\end{figure*}

We utilised the periods and extinction corrected intensity averaged mean magnitudes obtained in Section \ref{sec:RR Lyrae stars:template_fitting} to derive PLRs for the $R$ and $I$ bands. The PLRs are fitted to the data, assuming the mean magnitudes vary linearly with the $\log$ period. The following equation is fitted to the data,

\begin{equation}\label{eq:PLR}
    {\rm m_{\lambda}} = {\rm b_{\lambda} \log(P) +  a_{\lambda}},
\end{equation}
Here, ${\rm a_{\lambda}}$ and ${\rm b_{\lambda}}$ are the zero point and slope of the PLR in a given filter $\lambda$. The scatter (rms) in the PLRs is a consequence of the intrinsic width of the instability strip in temperature \citep{marconi_new_2015}. It may also arise due to the metallicity spread of the RR Lyrae in clusters and the uncertainties in extinction correction. However, M3 stars do not show a significant metallicity spread ($\sigma_{\feh} \sim 0.03 $ dex, \citealt{sneden_chemical_2004}). Fig. \ref{fig:PLR_PIR} represents the distribution of mean magnitudes with the logarithmic period in the $I$ and $R$ bands. In the left panel of plot, we show the $\log (P)$ vs mean magnitude in the respective band for RRab and RRc stars and in the right panel, the global PL relation is shown for I and R bands. An iterative linear fit is performed by removing the outlier stars ($ \geq 3\sigma $) to get the slope and intercept of the PLR for both RRab and RRc stars. Global PLRs were also derived by fundamentalising the periods of RRc variables using the relation \citep{iben_comments_1971, rood_metal-poor_1973, cox_double-mode_1983, di_criscienzo_rr_2004, coppola_carina_2015},
\begin{equation}\label{eq:fundamentalisation}
\log{P_{\rm FU}} = \log{P_{\rm FO}} + 0.127,
\end{equation}
Where `FU' refers to the fundamental mode and `FO' represents the first-overtone mode.

Table \ref{tab:PLR} lists the values of slopes and zero points of the Period-Luminosity relations for RRc and RRab stars as well as global relations. These empirical relations are best constrained in the $I$ band with a scatter of only 0.149 mag for RRc stars, 0.107 mag for RRab, and 0.124 mag for the combined sample of RRab and RRc stars.

\subsection{Period-Wesenheit (PW) Relations}\label{sec:PW_relations}
The Wesenheit index is defined as a combination of multiple passband magnitudes of a star in a way that effectively eliminates the impact of reddening. The Wesenheit index was initially developed to study Cepheid variable stars' Period-Luminosity relations \citep{madore_Period-Luminosity_1982}. Over time, the application of the Wesenheit index has expanded beyond Cepheids and has been used in various studies of different types of variable stars, including RR Lyrae stars, which are crucial for distance determinations in globular clusters and studies of stellar populations in galaxies \citep{neeley_standard_2019, mullen_rr_2023}.

\begin{figure*}
    \centering
    \includegraphics[]{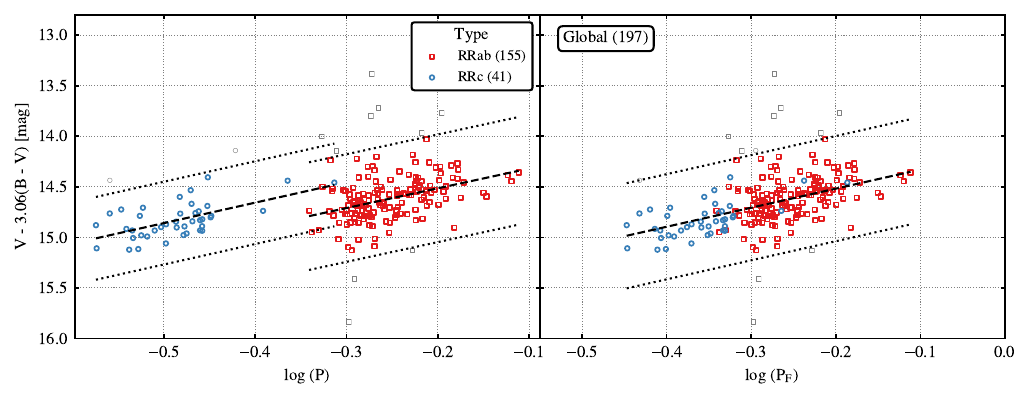} \\
    \includegraphics[]{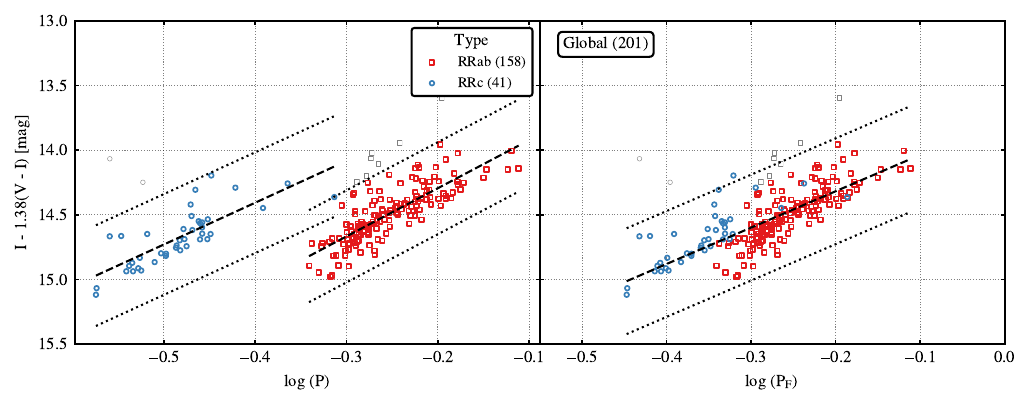}
    \caption{Empirical dual band Period-Wesenheit relations for M3 RR Lyrae stars. The fitted PW relation is represented by the dashed line, and the region of $\pm 3 \sigma$ is represented by the dotted lines.}
    \label{fig:PW}
\end{figure*}

\begin{table*}
    \centering
    \caption{This table presents the coefficients (slopes and zero points) for the empirical Period-Wesenheit relations of M3. The relationship is described by the equation W($X_1$, $X_2 - X_3$) = M$_{X_1}$ - $\zeta$(M$_{X_2}$ - M$_{X_3}$) = b $\log ({\rm P})$ + a, where $X_1$, $X_2$, and $X_3$ represent the filter bands, and M$_{X_1}$, M$_{X_2}$, and M$_{X_3}$ denote the mean magnitudes in the corresponding filters used to calculate the Wesenheit index. The parameter $\zeta$ is adopted from \citet{marconi_new_2015}, and `N' indicates the number of stars included in the fit. }
    \label{tab:PW_fitting}
    \resizebox{\linewidth}{!}{%
\begin{tabular}{lrcccccccccccc}
\hline
PW         & $\zeta$  & \multicolumn{4}{c}{RRab} & \multicolumn{4}{c}{RRc} & \multicolumn{4}{c}{Global} \\
         &      & N    &        b &                   a &    $\sigma$ & N &                   b &                   a &    $\sigma$ & N &                   b &                   a &    $\sigma$ \\
\hline
 & \multicolumn{12}{c}{Dual band PW relations}  &  \\
\hline
V, (B-V) & 3.06 & 155 & -1.955 $\pm$ 0.322 & 14.124 $\pm$ 0.082 & 0.178 & 41 & -2.026 $\pm$ 0.427 & 13.846 $\pm$ 0.208 & 0.136 & 197 & -1.885 $\pm$ 0.194 & 14.141 $\pm$ 0.055 & 0.173 \\
I, (B-I) & 0.78 & 152 & -3.190 $\pm$ 0.145 & 13.719 $\pm$ 0.037 & 0.080 & 41 & -3.027 $\pm$ 0.369 & 13.247 $\pm$ 0.179 & 0.119 & 190 & -2.750 $\pm$ 0.106 & 13.815 $\pm$ 0.030 & 0.089 \\
I, (V-I) & 1.38 & 158 & -3.724 $\pm$ 0.211 & 13.550 $\pm$ 0.054 & 0.119 & 41 & -3.231 $\pm$ 0.401 & 13.115 $\pm$ 0.195 & 0.130 & 201 & -2.812 $\pm$ 0.156 & 13.756 $\pm$ 0.044 & 0.136 \\
J, (B-J) & 0.27 & 156 & -2.523 $\pm$ 0.089 & 13.932 $\pm$ 0.023 & 0.049 & 43 & -2.536 $\pm$ 0.160 & 13.606 $\pm$ 0.078 & 0.054 & 199 & -2.527 $\pm$ 0.056 & 13.931 $\pm$ 0.016 & 0.050 \\
J, (V-J) & 0.40 & 159 & -2.656 $\pm$ 0.108 & 13.890 $\pm$ 0.028 & 0.060 & 43 & -2.596 $\pm$ 0.186 & 13.576 $\pm$ 0.091 & 0.062 & 202 & -2.622 $\pm$ 0.068 & 13.898 $\pm$ 0.019 & 0.061 \\
J, (I-J) & 0.96 & 159 & -2.254 $\pm$ 0.178 & 14.026 $\pm$ 0.046 & 0.100 & 43 & -2.622 $\pm$ 0.375 & 13.641 $\pm$ 0.183 & 0.125 & 201 & -2.539 $\pm$ 0.120 & 13.962 $\pm$ 0.034 & 0.105 \\
H, (B-H) & 0.16 & 158 & -2.739 $\pm$ 0.066 & 13.711 $\pm$ 0.017 & 0.037 & 43 & -2.861 $\pm$ 0.118 & 13.298 $\pm$ 0.057 & 0.039 & 201 & -2.733 $\pm$ 0.042 & 13.711 $\pm$ 0.012 & 0.038 \\
H, (V-H) & 0.22 & 156 & -2.797 $\pm$ 0.067 & 13.688 $\pm$ 0.017 & 0.038 & 43 & -2.902 $\pm$ 0.126 & 13.274 $\pm$ 0.061 & 0.042 & 199 & -2.778 $\pm$ 0.043 & 13.692 $\pm$ 0.012 & 0.039 \\
H, (I-H) & 0.44 & 160 & -2.666 $\pm$ 0.092 & 13.710 $\pm$ 0.024 & 0.051 & 42 & -2.832 $\pm$ 0.156 & 13.313 $\pm$ 0.076 & 0.051 & 202 & -2.787 $\pm$ 0.058 & 13.681 $\pm$ 0.016 & 0.052 \\
H, (J-H) & 1.68 & 156 & -2.912 $\pm$ 0.098 & 13.458 $\pm$ 0.025 & 0.054 & 41 & -3.148 $\pm$ 0.181 & 12.985 $\pm$ 0.089 & 0.060 & 198 & -3.013 $\pm$ 0.064 & 13.432 $\pm$ 0.018 & 0.057 \\
Ks, (B-Ks) & 0.10 & 161 & -2.582 $\pm$ 0.077 & 13.785 $\pm$ 0.020 & 0.043 & 43 & -2.793 $\pm$ 0.134 & 13.359 $\pm$ 0.065 & 0.045 & 204 & -2.631 $\pm$ 0.049 & 13.772 $\pm$ 0.014 & 0.044 \\
Ks, (V-Ks) & 0.13 & 161 & -2.597 $\pm$ 0.079 & 13.783 $\pm$ 0.020 & 0.044 & 43 & -2.809 $\pm$ 0.141 & 13.353 $\pm$ 0.069 & 0.047 & 204 & -2.640 $\pm$ 0.050 & 13.771 $\pm$ 0.014 & 0.045 \\
Ks, (I-Ks) & 0.25 & 163 & -2.528 $\pm$ 0.099 & 13.793 $\pm$ 0.025 & 0.056 & 43 & -2.842 $\pm$ 0.169 & 13.343 $\pm$ 0.083 & 0.057 & 207 & -2.691 $\pm$ 0.064 & 13.752 $\pm$ 0.018 & 0.058 \\
Ks, (J-Ks) & 0.69 & 162 & -2.623 $\pm$ 0.091 & 13.710 $\pm$ 0.023 & 0.051 & 43 & -2.919 $\pm$ 0.177 & 13.238 $\pm$ 0.087 & 0.059 & 204 & -2.700 $\pm$ 0.058 & 13.691 $\pm$ 0.016 & 0.052 \\
Ks, (H-Ks) & 1.87 & 162 & -2.406 $\pm$ 0.146 & 13.884 $\pm$ 0.038 & 0.083 & 43 & -2.679 $\pm$ 0.248 & 13.462 $\pm$ 0.121 & 0.083 & 205 & -2.514 $\pm$ 0.092 & 13.858 $\pm$ 0.026 & 0.083 \\

\hline
 & \multicolumn{12}{c}{Triple band PW relations}  &  \\
\hline
V, (B-I) & 1.34 & 155 & -2.949 $\pm$ 0.170 & 13.801 $\pm$ 0.044 & 0.095 & 42 & -2.733 $\pm$ 0.408 & 13.407 $\pm$ 0.198 & 0.133 & 197 & -2.429 $\pm$ 0.121 & 13.918 $\pm$ 0.034 & 0.106 \\
J, (B-I) & 0.38 & 155 & -2.644 $\pm$ 0.080 & 13.891 $\pm$ 0.021 & 0.045 & 42 & -2.638 $\pm$ 0.148 & 13.529 $\pm$ 0.072 & 0.048 & 199 & -2.499 $\pm$ 0.055 & 13.925 $\pm$ 0.015 & 0.048 \\
J, (V-I) & 0.68 & 158 & -2.848 $\pm$ 0.089 & 13.819 $\pm$ 0.023 & 0.050 & 42 & -2.728 $\pm$ 0.170 & 13.465 $\pm$ 0.083 & 0.056 & 200 & -2.609 $\pm$ 0.060 & 13.874 $\pm$ 0.017 & 0.053 \\
H, (B-I) & 0.24 & 155 & -2.788 $\pm$ 0.058 & 13.719 $\pm$ 0.015 & 0.033 & 43 & -2.790 $\pm$ 0.134 & 13.336 $\pm$ 0.065 & 0.045 & 199 & -2.666 $\pm$ 0.042 & 13.747 $\pm$ 0.012 & 0.037 \\
H, (V-I) & 0.42 & 155 & -2.900 $\pm$ 0.063 & 13.683 $\pm$ 0.016 & 0.035 & 43 & -2.827 $\pm$ 0.151 & 13.308 $\pm$ 0.074 & 0.050 & 199 & -2.723 $\pm$ 0.046 & 13.723 $\pm$ 0.013 & 0.041 \\
Ks, (B-I) & 0.16 & 160 & -2.623 $\pm$ 0.070 & 13.782 $\pm$ 0.018 & 0.039 & 43 & -2.755 $\pm$ 0.131 & 13.376 $\pm$ 0.064 & 0.044 & 203 & -2.605 $\pm$ 0.045 & 13.785 $\pm$ 0.013 & 0.040 \\
Ks, (V-I) & 0.28 & 160 & -2.712 $\pm$ 0.069 & 13.755 $\pm$ 0.018 & 0.039 & 43 & -2.780 $\pm$ 0.139 & 13.357 $\pm$ 0.068 & 0.047 & 202 & -2.642 $\pm$ 0.045 & 13.770 $\pm$ 0.013 & 0.040 \\

\hline

\end{tabular}
}
\end{table*}



We adopted the same Wesenheit magnitudes constructed for theoretical models in \cite{marconi_new_2015}. We derive empirical Period-Wesenheit (PW) relations in the form of equation \ref{eq:PLR} following the same process described above for the PL Relations. We incorporated the mean magnitudes of M3 RR Lyrae stars in NIR bands (J, H, and K$_{\rm s}$) from \cite{bhardwaj_near-infrared_2020} to establish optical, optical-NIR double, and triple band period-wesenheit relations. Fig. \ref{fig:PW} shows dual band Wesenheit indices plotted against the log period. The plot shows the period vs Wesenheit indices separately for RRab and RRc stars, and the global sample of all RR Lyrae variables. Table \ref{tab:PW_fitting} contains the coefficients, their corresponding errors, and the standard deviations of the optical and optical-NIR dual and triple band PW relations for RR Lyrae stars in the M3 cluster.

\subsection{Distance to M3}\label{sec:distance_M3}

Using the derived PW relations for FU, FO, and the global sample of stars, we calculated the distances to each RR Lyrae star and then determined the mean distance to the M3 cluster. To do this, we employed the metal-independent PW($V$, $B-V$) relation provided by \cite{marconi_new_2015} as an absolute calibration and compared it with the observed PW($V$, $B-V$) relation. As the M3 globular cluster is a mono-metallic cluster with $\feh \sim -1.5$ dex \citep{harris_new_2010} and a small metallicity spread of $\Delta \feh \sim 0.03$ dex \citep{sneden_chemical_2004}, the effect of metallicity on distances derived using the PW relation is negligible. However it should be noted that the recent analysis by \cite{lee_multiple_2021} revealed a bimodal distributions in two populations, with $\langle \feh \rangle \approx -1.60$ and -1.45 dex. Since PW relations are independent of reddening, they are known for their accuracy in deriving distances of RR Lyrae stars \citep{braga_distance_2015, braga_rr_2016, bellinger_when_2020, kumar_predicting_2023}.

Through a comparison of the metal-independent theoretical calibrations by \cite{marconi_new_2015} with the observed slopes and zero-points, we obtained distance moduli ($\mu$) to M3 of 15.03 $\pm$ 0.03 (statistical) $\pm$ 0.12 (systematic) mag for FU, 15.05 $\pm$ 0.04 $\pm$ 0.30 mag for FO, and 15.05 $\pm$ 0.04 $\pm$ 0.08 mag for the global sample. The statistical error accounts for the dispersion in the distribution of individual RR Lyrae star distance moduli. In contrast, the systematic error reflects the discrepancy between the theoretical and semi-empirical calibration of the PW($V$, $B-V$) relations. Final estimates of the distance moduli is derived by taking average of $\mu_{\rm RRab}$, $\mu_{\rm RRc}$ and $\mu_{\rm Global}$. The resulting distance modulus using metal-independent theoretical calibrations is $\mu = 15.04 \pm 0.19 ${\rm {(syst.)}}$ \pm 0.04 {\rm  (stats)}$ mag. The estimates agree within 1$\sigma$ of the combined statistical and systematic errors for all samples of RR Lyrae stars.

Using the metal-dependent PW($I$, $V-I$) relation and adopting $\feh = -1.53$ dex, we derived the distance to M3. Employing the theoretical relations from \cite{marconi_new_2015}, we obtained the following distance moduli: 15.05 $\pm$ 0.06 $\pm$ 0.08 mag for FU stars, 15.01 $\pm$ 0.02 $\pm$ 0.28 mag for FO stars, and 15.05 $\pm$ 0.03 $\pm$ 0.06 mag for the global sample. The estimated distance modulus using metal-dependent theoretical calibrations is $\mu = 15.03 \pm 0.17 ${\rm {(syst.)}}$ \pm 0.04$ mag. Notably, both the metal-dependent and metal-independent estimates show agreement within the respective given standard errors. This outcome was expected, as the M3 cluster is characterised by being mono-metallic, with a minimal spread in its metallicity distribution. In a study of NIR photometry of RR Lyrae stars in M3, \cite{bhardwaj_near-infrared_2020} derived the distance to the M3 cluster using the theoretical predicted absolute calibrations of Period-Luminosity-metallicity relations in $JHK_s$ bands, yielding a distance modulus of $\mu = 15.041 \pm 0.017\,(\mathrm{stat.})\,\pm 0.036\,(\mathrm{syst.})$ mag. Our results based on optical photometry are also in excellent agreement with those based on near-infrared photometry \citep[e.g.,][]{bhardwaj_precise_2023}.

\section{Stellar parameters for RR Lyrae stars in M3}\label{sec:fourier_analysis_ANN}

\subsection{Fourier Analysis of RR Lyrae Stars in M3}\label{sec:fourier_analysis}

The shape of RR Lyrae stars' light curves contains useful information about their intrinsic properties.
Previous studies, such as those by \cite{simon_provisional_1993, kovacs_new_1995, jurcsik_determination_1996, kovacs_light_1996, kovacs_modelling_1998, deb_light_2009, bhardwaj_variation_2015, das_variation_2018} have investigated the correlation between Fourier parameters, as well as period, with intrinsic properties like mass, luminosity, metallicity, and colours.

In our study, we have an extensive collection of photometric data for M3, covering all optical bands. Leveraging this dataset, we derived the Fourier parameters for individual RR Lyrae light curves by fitting a Fourier series of sines to each light curve. Specifically, we applied a Fourier sine series to the light curves in filter bands with substantial phase coverage, namely the B, V, and I bands. The Fourier parameters were derived using the following equation:
\begin{equation}
    m = m_{0} + \sum_{k=1}^{N} A_{k} \sin(2k\pi\cdot x + \phi_k),
\end{equation}
For a specific filter band, $m$ represents the magnitude of the star at a given phase $x$, $m_0$ represents the mean magnitude, and $A_k$ and $\phi_k$ represent the Fourier amplitude and phase coefficients, respectively. Here, $N$ is order of fit, and we chose $N$ = 5, to fit the light curves.

We calculated the Fourier amplitude ratios ($R_{k1}$) and phase differences ($\phi_{k1}$) using the following equations:
\begin{equation}
\begin{split}
    R_{k1} &= \frac{A_k}{A_1}, \\
    \phi_{k1} &= \phi_k - k\phi_1,
\end{split}
\end{equation}
Here, $k$ is an integer greater than 1, and $0 \leq \phi_{k1} \leq 2\pi$. The errors associated with the Fourier parameters were calculated using error propagation methods applied to the Fourier coefficients. The Fourier amplitude ratios and phase differences for RR Lyrae in M3 along with the scatter ($\sigma$), are given in Table \ref{tab:FP_table} for the $BVI$ bands. 


\begin{table*}
    \centering
    \caption{The table presents the Fourier coefficients, along with their corresponding standard deviations, for the RR Lyrae stars in the M3 globular cluster, categorized according to the specified filter band. The complete table can be accessed online in a machine-readable format. }
    \resizebox{\textwidth}{!}{
    \begin{tabular}{llrclccccccccc}
\hline
Id & Type & Period & Band & N & $R_{{21}}$ & $\Phi_{{21}}$ & $R_{{31}}$ & $\Phi_{{31}}$ & $R_{{41}}$ & $\Phi_{{41}}$ & $R_{{51}}$ & $\Phi_{{51}}$ &$\sigma$ \\
\hline
V1 & RRab & 0.520590 & B & 271 & 0.415$\pm$0.012 & 1.996$\pm$0.032 & 0.316$\pm$0.012 & 4.561$\pm$0.046 & 0.204$\pm$0.011 & 0.831$\pm$0.065 & 0.183$\pm$0.012 & 3.294$\pm$0.075 & 0.085 \\
V1 & RRab & 0.520590 & V & 276 & 0.456$\pm$0.010 & 2.101$\pm$0.023 & 0.361$\pm$0.009 & 4.660$\pm$0.033 & 0.245$\pm$0.008 & 0.824$\pm$0.047 & 0.200$\pm$0.009 & 3.454$\pm$0.056 & 0.066 \\
V1 & RRab & 0.520590 & I & 75 & 0.476$\pm$0.026 & 2.521$\pm$0.061 & 0.395$\pm$0.019 & 5.197$\pm$0.092 & 0.229$\pm$0.018 & 1.760$\pm$0.139 & 0.142$\pm$0.018 & 4.462$\pm$0.175 & 0.055 \\
V3 & RRab & 0.558197 & B & 250 & 0.440$\pm$0.010 & 2.299$\pm$0.032 & 0.324$\pm$0.011 & 4.892$\pm$0.045 & 0.223$\pm$0.011 & 1.345$\pm$0.056 & 0.170$\pm$0.010 & 4.165$\pm$0.077 & 0.076 \\
\vdots & \vdots & \vdots & \vdots &    \vdots &       \vdots &   \vdots &     \vdots &         \vdots &    \vdots &\vdots &\vdots & \vdots & \vdots\\
V271 & RRab & 0.632800 & I & 35 & 0.464$\pm$0.026 & 3.051$\pm$0.079 & 0.224$\pm$0.024 & 5.908$\pm$0.151 & 0.138$\pm$0.028 & 2.899$\pm$0.202 & 0.086$\pm$0.027 & 6.018$\pm$0.319 & 0.035 \\
V292 & RRc & 0.296543 & B & 262 & 0.071$\pm$0.017 & 2.933$\pm$0.242 & 0.022$\pm$0.017 & 0.488$\pm$0.796 & 0.020$\pm$0.017 & 6.119$\pm$0.868 & 0.045$\pm$0.016 & 0.200$\pm$0.397 & 0.035 \\
V292 & RRc & 0.296543 & V & 276 & 0.099$\pm$0.022 & 3.522$\pm$0.196 & 0.057$\pm$0.021 & 2.126$\pm$0.369 & 0.067$\pm$0.023 & 2.175$\pm$0.290 & 0.042$\pm$0.021 & 3.466$\pm$0.484 & 0.042 \\
V292 & RRc & 0.296543 & I & 66 & 0.163$\pm$0.038 & 3.433$\pm$0.277 & 0.178$\pm$0.046 & 1.559$\pm$0.240 & 0.034$\pm$0.042 & 4.046$\pm$1.255 & 0.137$\pm$0.041 & 1.067$\pm$0.355 & 0.031 \\

\hline
\end{tabular}

    }
    \label{tab:FP_table}
\end{table*}

\begin{figure}
    \centering    
    \includegraphics[]{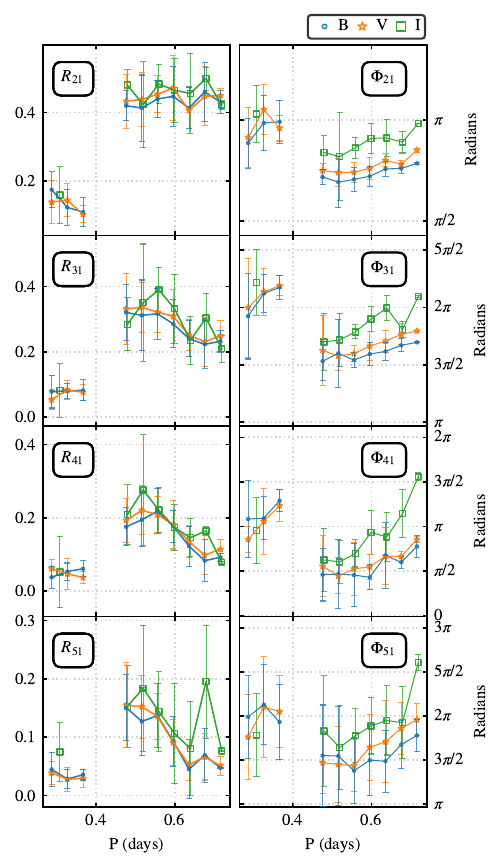}
    \caption{In this plot, the distribution of Fourier parameters is depicted alongside the period for RR Lyrae stars. Each data point represents the mean value of the Fourier parameter within a specific period bin, with error bars indicating the standard deviation in that corresponding period bin. The Fourier amplitude ratio parameters and the Fourier phase difference parameters are binned using a period bin size of 0.04 days.}
    \label{fig:P_vs_R_Phi}
\end{figure}

Fig. \ref{fig:P_vs_R_Phi} illustrates the relationship between the Fourier parameters ($R_{21}, R_{31}, \phi_{21}, \phi_{31}, ...$) and the period of RR Lyrae stars in M3 for each optical band light curve. The plot represents the mean Fourier parameters within each period bin, with the error bars indicating the standard deviation in that bin. The Fourier phase parameter ($\phi_{31}$), as discussed by \cite{jurcsik_determination_1996}, is an important parameter due to its dependence on the pulsation period and metallicity. We observe a linear relation between $\phi_{31}$ and the period, which is consistent with previous findings by \cite{jurcsik_photometric_2017} for $V$ band light curves. Moreover, Fourier phase parameters increase with wavelength, a trend that is seen for both RR Lyrae \citep{das_variation_2018} and Cepheid variables \citep{bhardwaj_variation_2015, bhardwaj_comparative_2017}.

\subsection{Physical parameters of non-Blazhko RR\MakeLowercase{ab} stars using Artificial Neural Network (ANN)}\label{sec:physical_parameters}

Physical parameters of RR Lyrae stars are usually derived using the empirical relations between various properties of a light curve like Fourier parameters, colour, etc., along with the period \citep{cacciari_multicolor_2005, deb_physical_2010, nemec_fourier_2011}. However, with the advancements in the theoretical modelling of stellar pulsation, it has been possible to generate a grid of models to study the properties of RR Lyrae and other variables stars \citep{marconi_new_2015, de_somma_extended_2020, de_somma_updated_2022}. The radial stellar pulsations (RSP) code of \cite{smolec_convective_2008} as a module in Modules for Experiments in Stellar Astrophysics \citep[\texttt{MESA},][]{paxton_modules_2011, paxton_modules_2013, paxton_modules_2015, paxton_modules_2018, paxton_modules_2019} is also available and can be used to generate new theoretical models. Such models are then used to constrain the physical properties of RR Lyrae stars, including their physical parameters like (mass, luminosity, effective temperature, etc).

The physical parameters of RR Lyrae can also be inferred by comparing the light curves of observed stars with a reference library of models. \cite{das_variation_2018} derived the physical parameters of a few RRab stars in the large magellanic cloud (LMC) by comparing the observed light curves with the theoretical model light curves provided by \cite{marconi_new_2015}. A drawback of the method is that only a small number of LMC light curves could be matched with theoretical models.

Another approach to derive the physical parameters using theoretical models is to use a denser and smoother grid of models with non-linear optimisation methods \citep{bellinger_fundamental_2016}. However, the models require a high amount of computational resources for computing even just a single light curve, and it is thus not feasible to generate a smooth and dense grid. To overcome this problem, \cite{kumar_predicting_2023} trained an artificial neural network (ANN) on the physical parameter - light curve model grid of \cite{marconi_new_2015} for RRab stars in $V$ and $I$ bands. The trained neural network\footnote{The trained ANN networks can be accessed online on \url{https://ann-interpolator.web.app/}.} can generate a much smoother and denser grid of theoretical models, and one light curve takes only $\sim$ 55 ms to generate. 

Taking into account the predicted evolutionary properties of RR Lyrae for the adopted metal abundance (Z=0.001 which corresponds to $\feh$=-1.54 dex), we used the ANN interpolator to generate a grid of RRab models with M = 0.58-0.64 $\msun$, with a step size of 0.003 $\msun$,  $\log( {\rm L}/ {\rm L}_{\odot})$ = 1.67 - 1.99 dex, with a step size of 0.016 dex, and effective temperature ranging from 5700-6900 K with a step size of 60 K. We fixed the hydrogen abundance ratio, X to $0.754$. The selection of limits for the stellar parameters was guided by the boundaries defined within the parameter space of models of \cite{marconi_new_2015}. This choice aligns with the fact that the ANN is trained exclusively on these models and may struggle to generate accurate light curves for parameters lying beyond the boundaries of the original grid. We generated $8000$ new $V$ band light curves with the given combination of mass, luminosity, and temperature.

To derive the physical parameters of the observed RRab stars in the M3 cluster, we generated a grid of models and compared them to the observed $V$ band light curves because they have the best sampling. The relationship between the light curve ($y$) and the corresponding physical parameters ($\textbf{x}$) is represented by a function ($f$), such that $y \equiv f(\textbf{x})$. Conversely, an inverse function ($g$) exists, enabling the representation of $\textbf{x}$ in terms of $y$, i.e., $\textbf{x} \equiv f^{-1}(y) \equiv g(y)$. Artificial Neural Networks (ANNs) serve as effective function approximators when the functions are continuous and differentiable \citep{cybenko_approximation_1989, hornik_multilayer_1989, hornik_approximation_1991}. 

In this study, an ANN called \emph{RRab-Net} is trained using 8000 newly generated models to approximate the function `$g$'. The selection of architecture and hyperparameters in our study was typically based on expert intuition and manual tuning. We utilised a trial-and-error approach, random grid search \citep{bergstra_random_2012}, to determine these characteristics. A grid of possible hyperparameter combinations is tabulated in Table~\ref{tab:hyperparameter_combinations}. From these combinations, we randomly selected a set of $200$ hyperparameter combinations. The network was then trained for a fixed $1000$ epochs using the L2 norm (MSE) as the objective function and the adaptive moment stochastic gradient descent (or \texttt{adam}: \citealt{kingma_adam_2014} algorithm with a default batch size of $32$ samples) as an optimization algorithm. The optimisation of the network architecture was performed using the \texttt{KerasTuner}\footnote{\url{https://keras.io/keras_tuner/}.} \citep[][]{omalley_kerastuner_2019} module of Python. The final network architecture adopted for the training of \emph{RRab-net} consists of three hidden layers with 128, 64 and 64 neurons, respectively, in those layers, as detailed in Table \ref{tab:network_architecture}. The details of the final training of \emph{RRab-Net} are discussed in Appendix \ref{sec:apndx_train_ANN}.

\begin{table}
    \centering
    \caption{The hyperparameter search space for the artificial neural network.}
    \resizebox{\linewidth}{!}{%
    \begin{tabular}{cll}
        \hline 
        S. N. & Name of hyperparameter & Possible values  \\
        \hline 
        1 & Number of hidden layers & [1, 2, 3] \\
        2 & Number of neurons in  one hidden layer & [16, 32, 64, 128]   \\
        3 & Optimizer & `\texttt{adam}' \citep{kingma_adam_2014} \\
        4 & Learning rate  & [$10^{-2} - 10^{-4}$](log sampling) \\
        5 & Activation function & [`\texttt{relu}', `\texttt{tanh}'] \\
        6 & Weights initialisation & `\texttt{GlorotUniform}' \\
        \hline
    \end{tabular}
    }
    \label{tab:hyperparameter_combinations}
\end{table}

The accuracy of the predicted physical parameters is assessed by comparing the observed light curves with those generated by the ANN interpolator, a neural network trained by \cite{kumar_predicting_2023}. If the generated light curve closely matches the observed light curve, it validates the accuracy of the inferred physical parameters. The mean squared error (MSE) and correlation coefficient (R$^{2}$) are utilised as metrics to assess the similarity between the predicted and observed light curves. Based on the metrics employed, our findings indicate that the artificial neural network (ANN) can successfully predict light curves that closely resemble the observed light curves for many stars. However, it is essential to note instances where the predicted and observed light curves do not align. This can be attributed to certain limitations in our approach. As our initial network is trained on a finite grid of models, it might not encompass the full range of possible physical parameter combinations. Consequently, when dealing with observed light curves associated with physical parameters falling beyond the scope of our training grid, our method might yield less accurate estimations of these parameters.

\begin{figure}
    \centering
    \includegraphics{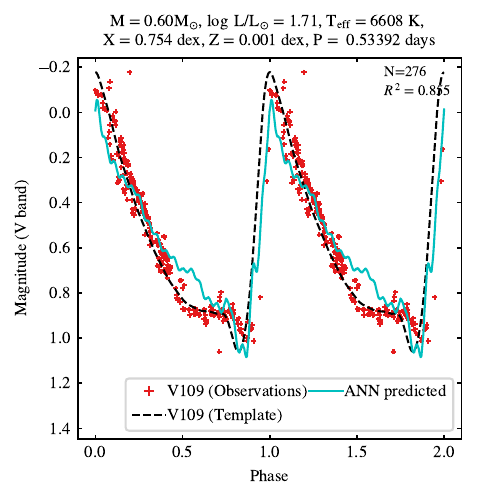}
    \caption{This figure displays the observed and ANN-predicted light curve of V109. The ANN-predicted light curve is generated using the ANN interpolator introduced by \citet{kumar_predicting_2023}. The input parameters for the ANN interpolator are derived from \emph{RRab-Net}.}
    \label{fig:predicted_lc_vs_observed}
\end{figure}

Fig. \ref{fig:predicted_lc_vs_observed} illustrates the ANN-predicted light curve in the $V$ band, along with the inferred physical parameters for variable V109. The period, along with other physical parameters, is employed as input to generate the light curve. The observed light curve is scaled to absolute magnitude using the distance modulus derived in Section \ref{sec:distance_M3}. Note that the higher amplitude of theoretical light curves is a systematic in stellar pulsation models that can be mitigated by assuming a higher efficiency of convection \citep{bhardwaj_comparative_2017}.

\begin{figure}
    \centering
    \includegraphics{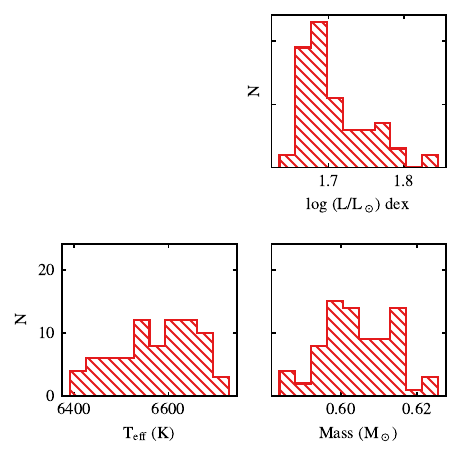}
    \caption{This figure presents histograms depicting the predicted mass, luminosity, and effective temperature derived from the $V$ band light curves with \emph{RRab-Net}.}
    \label{fig:MLT_hist}
\end{figure}

\begin{figure}
    \centering
    \includegraphics{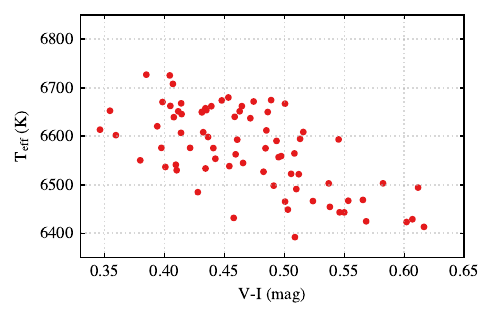}
    \caption{In this figure, the color ($V$-$I$) of stars is plotted against the derived effective temperature using RRab-Net.}
    \label{fig:Teff_VI}
\end{figure}

The masses (M/M$_\odot$), luminosities (expressed as $\log ({\rm L}/{\rm L}_{\odot})$), and effective temperatures ($\teff$) of non-Blazhko RRab stars have been deduced by analysing their $V$ band light curves. These parameters are provided in Table \ref{tab:predicted_params}, while their distributions are shown in Figure~\ref{fig:MLT_hist}. The average mass, luminosity, and effective temperature are $\rm{M} = 0.605 \pm 0.009 \rm{M}_\odot$, $\log( {\rm L}/ {\rm L}_{\odot}) = 1.71 \pm 0.04$ dex, and $\teff = 6571 \pm 83$ K. In Fig. \ref{fig:Teff_VI}, the derived effective temperatures are plotted against the color ($V$- $I$) of the non-Blazhko RRab stars and we found the expected correlation between them.

Several literature studies have explored the stellar parameters of non-Blazhko RRab stars in M3. \cite{cacciari_multicolor_2005} utilized colour-temperature calibrations to derive parameters for V72, yielding values of M = $0.72 \pm 0.05 \msun$, $\log( {\rm L}/ {\rm L}_{\odot}) = 1.67 \pm 0.03 $, and $\teff = 6773 \pm 100 $ K. \cite{marconi_modeling_2007} conducted a comprehensive analysis of M3 RR Lyrae stars, including V128, V126, V72, and V152. However, only V72 is a non-Blazhko RRab star among these RR Lyrae variables. They determined the following properties for this star: M ranging from $0.60$ to $0.73 \msun$, $\log ({\rm L}/{\rm L}_{\odot})$ spanning between 1.643 and 1.710 dex, and $\teff$ ranging from 6900 to 7000 K. However, in our study, focusing on the same star, we obtained slightly different values: M = $0.60 \msun $, $\log( {\rm L}/ {\rm L}_{\odot}) = 1.67$, and $\teff = 6563$ K. In a related study, \cite{valcarce_a_a_r_semi-empirical_2008} investigated the mass distribution of horizontal branch stars within M3. Employing a semi-empirical approach, they determined a mass distribution for non-Blazhko RRab stars, yielding a value of 0.644 $\pm$ 0.005 $\msun$ for the masses of those stars. In contrast, our analysis using $V$ band light curves predicted a mass distribution of $0.605 \pm 0.009$ $\msun$. These discrepancies emphasize the complexities and challenges inherent in determining precise physical parameters, likely arising from varying methodologies, data quality, and assumptions. Apart from the systematics and limitations of the pulsation models, the errors in derived parameters can also come from sparsely sampled observational light curves. More detailed investigations are needed in the methodologies for deriving stellar parameters based on light curve characteristics that can provide better constraints for the predictions of stellar pulsation and evolution models.

\begin{table}
    \centering
    \caption{This table presents the predicted parameters for Non-Blazhko RRab variables obtained through the application of \emph{RRab-Net} on their $V$ band light curves. In this table, $\sigma$ represents the Mean Squared Error (MSE), and $R^2$ signifies the correlation coefficient between the light curve predicted by the ANN interpolator and the template fitted to the original observations. The complete table can be accessed online in a machine-readable format.}
    \begin{tabular}{lccccr}
\hline
   Id   &      M/M$_\odot$ &    $\teff$ &      $\log({\rm L}/{\rm L}_{\odot})$ & $\sigma$ &  $R^2$ \\
      &                  &    (K)     &      (dex)                           &          &        \\      
\hline
V1 & 0.617 & 6674 & 1.696 & 0.018 & 0.864 \\
V6 & 0.612 & 6658 & 1.677 & 0.013 & 0.894 \\
V9 & 0.610 & 6545 & 1.686 & 0.010 & 0.900 \\
V11 & 0.623 & 6651 & 1.702 & 0.020 & 0.857 \\
$\vdots$ &  $\vdots$ & $\vdots$ &  $\vdots$ &    $\vdots$ &  $\vdots$ \\
V254 & 0.594 & 6432 & 1.765 & 0.013 & 0.721 \\
V257 & 0.601 & 6593 & 1.783 & 0.008 & 0.899 \\
V258 & 0.591 & 6413 & 1.717 & 0.013 & 0.689 \\
V262 & 0.602 & 6465 & 1.743 & 0.012 & 0.673 \\
\hline
\end{tabular}

    \label{tab:predicted_params}
\end{table}

\section{Summary}\label{sec:summary}
This study on the M3 globular cluster encompassed various aspects to gain insights into the pulsation properties of its RR Lyrae population. We utilised a large photometric dataset spanning 35 years to obtain light curves of RR Lyrae stars at multiple filter bands. We recovered 238 previously known RR Lyrae stars in the M3 globular cluster, including 178 RRab stars, 49 RRc stars, and 11 RRd stars. Applying the Multi-Band Lomb Scargle (MBLS) algorithm \citep{vanderplas_periodograms_2015} on our long-term photometric time-series data improved the accuracy and precision of the periods of 154 stars. To obtain accurate measurements of mean magnitudes and amplitudes from light curves with limited phase coverage or containing erroneous data points, we utilised a template fitting at multiple wavelengths. 

Multi-band photometric data were used to investigate the observed topology of the instability strip on the colour-magnitude diagrams. Most RR Lyrae stars in M3 fall between the predicted boundaries of the instability strip based on stellar pulsation models. We filtered out outlier stars present in the dataset using optical-NIR CMDs. Stars that were outliers in over 75\% of the CMDs were rejected and not considered further. Bailey's diagram enabled a separation between RRab and RRc stars and confirmation of their classification. We also studied the relation of luminosity amplitude ratios (relative to the $V$ band) with the pulsation period. We observed a general agreement between the derived values of ratios in different bands and those reported in the literature. However, in contrast to the findings of \cite{bhardwaj_near-infrared_2020} for $JHK_s$ bands, we did not observe distinct values for long-period RRab stars. 

The Period-Luminosity relations for the $I$ and $R$ bands were derived for RRab and RRc stars. We also investigated the Period-Wesenheit relations for RR Lyrae stars in optical as well as NIR bands. The slopes for the FU and FO mode relation were found within 10\% and 15\% respectively when compared with theoretically predicted Period-Luminosity-metallicity relations \citep{marconi_new_2015}, while a slight discrepancy is observed for the Global mode. Using a combined sample of RR Lyrae stars, we derived a distance modulus of $\mu = 15.04 \pm 0.19 ${\rm {(syst.)}}$ \pm 0.04 {\rm  (stats)}$ mag using the metal-independent $W_{BV}$ Wesenheit magnitude. Similarly, using metal-dependent $W_{VI}$, Wesenheits were used to derive a distance modulus of $\mu = 15.03 \pm 0.17 ${\rm {(syst.)}}$ \pm 0.04$ mag. These results are in agreement with the current literature on the distance to the M3 cluster.


We studied the light curve structure of RR Lyrae stars in M3 using the Fourier decomposition method. The Fourier phase parameter $\phi_{31}$ exhibits a linear relation with the period, consistent with previous findings. Moreover, average phase parameters increase with wavelength at a given period. Finally, we utilised an artificial neural network that was trained with $8000$ newly generated $V$ band light curve models to determine the physical parameters of non-Blazhko RRab stars in M3. We then applied the neural network to the observed $V$ band light curves of RRab stars in M3, determining their physical parameters. For the non-Blazhko RRab stars, the average values of the physical parameters are M = $0.605 \pm 0.009$ M$\odot$, $\log( {\rm L}/ {\rm L}{\odot}) = 1.71 \pm 0.04 {\rm (stats)}$ dex, and $\teff = 6571 \pm 83$ K. These physical parameter estimates are within the diverse range of limited measurements for M3 RR Lyrae stars in the literature \citep{cacciari_multicolor_2005, marconi_modeling_2007, valcarce_a_a_r_semi-empirical_2008}. Determining stellar parameters accurately is a challenging task, and the empirical differences highlight the need to improve the accuracy of physical parameters. This can also be done by adopting a multi-wavelength approach, where the neural network is trained on light curves in several filters. More advanced machine-learning methods, together with the multi-wavelength data, will be explored in future to determine accurate and precise physical parameters of RR Lyrae stars, gaining new insights into our understanding of stellar evolution and pulsation.



\section*{Acknowledgements}
We are grateful to P. B. Stetson for generously sharing the optical light curves of RR Lyrae stars in M3 used in this study. The authors acknowledge the usage of the IUCAA HPC computing facility for numerical calculations (training ANN models). NK acknowledges the financial assistance from the Council of Scientific and Industrial Research (CSIR), New Delhi, India, as the Senior Research Fellowship (SRF) file no. 09/45(1651)/2019-EMR-I. AB acknowledges funding from the European Union’s Horizon 2020 research and innovation programme under the Marie Sk\l{}odowska-Curie grant agreement Number 886298. HPS acknowledges a grant from the Council of Scientific and Industrial Research (CSIR), New Delhi, India, file no. 03(1428)/18-EMR-II. 

\section*{Data Availability}
All the tables will be made available through Vizier.
 



\bibliographystyle{mnras}
\bibliography{paper} 

\begin{thebibliography}{}
\makeatletter
\relax
\def\mn@urlcharsother{\let\do\@makeother \do\$\do\&\do\#\do\^\do\_\do\%\do\~}
\def\mn@doi{\begingroup\mn@urlcharsother \@ifnextchar [ {\mn@doi@}
  {\mn@doi@[]}}
\def\mn@doi@[#1]#2{\def\@tempa{#1}\ifx\@tempa\@empty \href
  {http://dx.doi.org/#2} {doi:#2}\else \href {http://dx.doi.org/#2} {#1}\fi
  \endgroup}
\def\mn@eprint#1#2{\mn@eprint@#1:#2::\@nil}
\def\mn@eprint@arXiv#1{\href {http://arxiv.org/abs/#1} {{\tt arXiv:#1}}}
\def\mn@eprint@dblp#1{\href {http://dblp.uni-trier.de/rec/bibtex/#1.xml}
  {dblp:#1}}
\def\mn@eprint@#1:#2:#3:#4\@nil{\def\@tempa {#1}\def\@tempb {#2}\def\@tempc
  {#3}\ifx \@tempc \@empty \let \@tempc \@tempb \let \@tempb \@tempa \fi \ifx
  \@tempb \@empty \def\@tempb {arXiv}\fi \@ifundefined
  {mn@eprint@\@tempb}{\@tempb:\@tempc}{\expandafter \expandafter \csname
  mn@eprint@\@tempb\endcsname \expandafter{\@tempc}}}

\bibitem[\protect\citeauthoryear{Bailey}{Bailey}{1902}]{bailey_discussion_1902}
Bailey S.~I.,  1902, Annals of Harvard College Observatory, 38, 1

\bibitem[\protect\citeauthoryear{Beaton et~al.,}{Beaton
  et~al.}{2016}]{beaton_carnegie-chicago_2016}
Beaton R.~L.,  et~al., 2016, \mn@doi [\apj] {10.3847/0004-637X/832/2/210}, 832,
  210

\bibitem[\protect\citeauthoryear{Bellinger, Angelou, Hekker, Basu, Ball  \&
  Guggenberger}{Bellinger et~al.}{2016}]{bellinger_fundamental_2016}
Bellinger E.~P.,  Angelou G.~C.,  Hekker S.,  Basu S.,  Ball W.~H.,
  Guggenberger E.,  2016, \mn@doi [The Astrophysical Journal]
  {10.3847/0004-637X/830/1/31}, 830, 31

\bibitem[\protect\citeauthoryear{Bellinger, Kanbur, Bhardwaj  \&
  Marconi}{Bellinger et~al.}{2020}]{bellinger_when_2020}
Bellinger E.~P.,  Kanbur S.~M.,  Bhardwaj A.,   Marconi M.,  2020, \mn@doi
  [\mnras] {10.1093/mnras/stz3292}, 491, 4752

\bibitem[\protect\citeauthoryear{Benk\H{o}, Szabó  \& Paparó}{Benk\H{o}
  et~al.}{2011}]{benkho_blazhko_2011}
Benk\H{o} J.~M.,  Szabó R.,   Paparó M.,  2011, \mn@doi [\mnras]
  {10.1111/j.1365-2966.2011.19313.x}, 417, 974

\bibitem[\protect\citeauthoryear{Benkő, Bakos  \& Nuspl}{Benkő
  et~al.}{2006}]{benko_multicolour_2006}
Benkő J.~M.,  Bakos G.~A.,   Nuspl J.,  2006, \mn@doi [\mnras]
  {10.1111/j.1365-2966.2006.10953.x}, 372, 1657

\bibitem[\protect\citeauthoryear{Bergstra \& Bengio}{Bergstra \&
  Bengio}{2012}]{bergstra_random_2012}
Bergstra J.,  Bengio Y.,  2012, Journal of machine learning research, 13

\bibitem[\protect\citeauthoryear{Bhardwaj}{Bhardwaj}{2020}]{bhardwaj_high-precision_2020}
Bhardwaj A.,  2020, \mn@doi [Journal of Astrophysics and Astronomy]
  {10.1007/s12036-020-09640-z}, 41, 23

\bibitem[\protect\citeauthoryear{Bhardwaj}{Bhardwaj}{2022}]{bhardwaj_rr_2022}
Bhardwaj A.,  2022, Universe, 8, 122

\bibitem[\protect\citeauthoryear{Bhardwaj, Kanbur, Singh, Macri  \&
  Ngeow}{Bhardwaj et~al.}{2015}]{bhardwaj_variation_2015}
Bhardwaj A.,  Kanbur S.~M.,  Singh H.~P.,  Macri L.~M.,   Ngeow C.-C.,  2015,
  \mn@doi [\mnras] {10.1093/mnras/stu2678}, 447, 3342

\bibitem[\protect\citeauthoryear{Bhardwaj, Kanbur, Marconi, Rejkuba, Singh  \&
  Ngeow}{Bhardwaj et~al.}{2017}]{bhardwaj_comparative_2017}
Bhardwaj A.,  Kanbur S.~M.,  Marconi M.,  Rejkuba M.,  Singh H.~P.,   Ngeow
  C.-C.,  2017, \mn@doi [\mnras] {10.1093/mnras/stw3256}, 466, 2805

\bibitem[\protect\citeauthoryear{Bhardwaj, Rejkuba, de Grijs, Herczeg, Singh,
  Kanbur  \& Ngeow}{Bhardwaj et~al.}{2020}]{bhardwaj_near-infrared_2020}
Bhardwaj A.,  Rejkuba M.,  de Grijs R.,  Herczeg G.~J.,  Singh H.~P.,  Kanbur
  S.,   Ngeow C.-C.,  2020, \mn@doi [\aj] {10.3847/1538-3881/abb3f9}, 160, 220

\bibitem[\protect\citeauthoryear{Bhardwaj et~al.,}{Bhardwaj
  et~al.}{2021}]{bhardwaj_rr_2021}
Bhardwaj A.,  et~al., 2021, \mn@doi [\aj] {10.3847/1538-4357/abdf48}, 909, 200

\bibitem[\protect\citeauthoryear{Bhardwaj et~al.,}{Bhardwaj
  et~al.}{2023}]{bhardwaj_precise_2023}
Bhardwaj A.,  et~al., 2023, \mn@doi [The Astrophysical Journal Letters]
  {10.3847/2041-8213/acba7f}, 944, L51

\bibitem[\protect\citeauthoryear{Blazhko}{Blazhko}{1907}]{blazhko_mitteilung_1907}
Blazhko S.,  1907, Astronomische Nachrichten, 175, 325

\bibitem[\protect\citeauthoryear{Bono, Caputo, Castellani, Marconi  \&
  Storm}{Bono et~al.}{2001}]{bono_theoretical_2001}
Bono G.,  Caputo F.,  Castellani V.,  Marconi M.,   Storm J.,  2001, \mn@doi
  [\mnras] {10.1046/j.1365-8711.2001.04655.x}, 326, 1183

\bibitem[\protect\citeauthoryear{Braga et~al.,}{Braga
  et~al.}{2015}]{braga_distance_2015}
Braga V.~F.,  et~al., 2015, \mn@doi [\apj] {10.1088/0004-637X/799/2/165}, 799,
  165

\bibitem[\protect\citeauthoryear{Braga et~al.,}{Braga
  et~al.}{2016}]{braga_rr_2016}
Braga V.~F.,  et~al., 2016, The Astronomical Journal, 152, 170

\bibitem[\protect\citeauthoryear{Cacciari, Corwin  \& Carney}{Cacciari
  et~al.}{2005}]{cacciari_multicolor_2005}
Cacciari C.,  Corwin T.~M.,   Carney B.~W.,  2005, \mn@doi [\aj]
  {10.1086/426325}, 129, 267

\bibitem[\protect\citeauthoryear{Caputo}{Caputo}{1990}]{caputo_oosterhoff_1990}
Caputo F.,  1990, \aap, 239, 137

\bibitem[\protect\citeauthoryear{Castellani \& Quarta}{Castellani \&
  Quarta}{1987}]{castellani_oosterhoff_1987}
Castellani V.,  Quarta M.~L.,  1987, \aaps, 71, 1

\bibitem[\protect\citeauthoryear{Castellani, Castellani  \& Cassisi}{Castellani
  et~al.}{2005}]{castellani_rr_2005}
Castellani M.,  Castellani V.,   Cassisi S.,  2005, \mn@doi [\aap]
  {10.1051/0004-6361:20042306}, 437, 1017

\bibitem[\protect\citeauthoryear{Catelan}{Catelan}{2004a}]{catelan_rr_2004}
Catelan M.,  2004a, in Kurtz D.~W.,  Pollard K.~R.,  eds,  Astronomical
  {Society} of the {Pacific} {Conference} {Series} Vol. 310, {IAU} {Colloq}.
  193: {Variable} {Stars} in the {Local} {Group}. p.~113,
  \mn@doi{10.48550/arXiv.astro-ph/0310159}

\bibitem[\protect\citeauthoryear{Catelan}{Catelan}{2004b}]{catelan_evolutionary_2004}
Catelan M.,  2004b, \mn@doi [\apj] {10.1086/379657}, 600, 409

\bibitem[\protect\citeauthoryear{Catelan}{Catelan}{2009}]{catelan_horizontal_2009}
Catelan M.,  2009, \mn@doi [\apss] {10.1007/s10509-009-9987-8}, 320, 261

\bibitem[\protect\citeauthoryear{Catelan, Pritzl  \& Smith}{Catelan
  et~al.}{2004}]{catelan_rr_2004-1}
Catelan M.,  Pritzl B.~J.,   Smith H.~A.,  2004, \mn@doi [The Astrophysical
  Journal Supplement Series] {10.1086/422916}, 154, 633

\bibitem[\protect\citeauthoryear{Clement et~al.,}{Clement
  et~al.}{2001}]{clement_variable_2001}
Clement C.~M.,  et~al., 2001, \mn@doi [The Astronomical Journal]
  {10.1086/323719}, 122, 2587

\bibitem[\protect\citeauthoryear{Clementini, Gratton, Bragaglia, Carretta,
  Di~Fabrizio  \& Maio}{Clementini et~al.}{2003}]{clementini_distance_2003}
Clementini G.,  Gratton R.,  Bragaglia A.,  Carretta E.,  Di~Fabrizio L.,
  Maio M.,  2003, \mn@doi [The Astronomical Journal] {10.1086/367773}, 125,
  1309

\bibitem[\protect\citeauthoryear{Cohen \& Meléndez}{Cohen \&
  Meléndez}{2005}]{cohen_abundances_2005}
Cohen J.~G.,  Meléndez J.,  2005, \mn@doi [The Astronomical Journal]
  {10.1086/426369}, 129, 303

\bibitem[\protect\citeauthoryear{Coppola et~al.,}{Coppola
  et~al.}{2015}]{coppola_carina_2015}
Coppola G.,  et~al., 2015, \mn@doi [\apj] {10.1088/0004-637X/814/1/71}, 814, 71

\bibitem[\protect\citeauthoryear{Corwin \& Carney}{Corwin \&
  Carney}{2001}]{corwin_bv_2001}
Corwin T.~M.,  Carney B.~W.,  2001, \mn@doi [The Astronomical Journal]
  {10.1086/323918}, 122, 3183

\bibitem[\protect\citeauthoryear{Cox, Hodson  \& Clancy}{Cox
  et~al.}{1983}]{cox_double-mode_1983}
Cox A.~N.,  Hodson S.~W.,   Clancy S.~P.,  1983, \mn@doi [\apj]
  {10.1086/160762}, 266, 94

\bibitem[\protect\citeauthoryear{Cybenko}{Cybenko}{1989}]{cybenko_approximation_1989}
Cybenko G.,  1989, \mn@doi [Mathematics of Control, Signals and Systems]
  {10.1007/BF02551274}, 2, 303

\bibitem[\protect\citeauthoryear{Das, Bhardwaj, Kanbur, Singh  \& Marconi}{Das
  et~al.}{2018}]{das_variation_2018}
Das S.,  Bhardwaj A.,  Kanbur S.~M.,  Singh H.~P.,   Marconi M.,  2018, \mn@doi
  [\mnras] {10.1093/mnras/sty2358}, 481, 2000

\bibitem[\protect\citeauthoryear{De~Somma, Marconi, Molinaro, Cignoni, Musella
  \& Ripepi}{De~Somma et~al.}{2020}]{de_somma_extended_2020}
De~Somma G.,  Marconi M.,  Molinaro R.,  Cignoni M.,  Musella I.,   Ripepi V.,
  2020, \mn@doi [\aj Supplement Series] {10.3847/1538-4365/ab7204}, 247, 30

\bibitem[\protect\citeauthoryear{De~Somma, Marconi, Molinaro, Ripepi, Leccia
  \& Musella}{De~Somma et~al.}{2022}]{de_somma_updated_2022}
De~Somma G.,  Marconi M.,  Molinaro R.,  Ripepi V.,  Leccia S.,   Musella I.,
  2022, \mn@doi [The Astrophysical Journal Supplement Series]
  {10.3847/1538-4365/ac7f3b}, 262, 25

\bibitem[\protect\citeauthoryear{Deb \& Singh}{Deb \&
  Singh}{2009}]{deb_light_2009}
Deb S.,  Singh H.~P.,  2009, Astronomy \& Astrophysics, 507, 1729

\bibitem[\protect\citeauthoryear{Deb \& Singh}{Deb \&
  Singh}{2010}]{deb_physical_2010}
Deb S.,  Singh H.~P.,  2010, \mn@doi [\mnras]
  {10.1111/j.1365-2966.2009.15927.x}, 402, 691

\bibitem[\protect\citeauthoryear{Denissenkov, VandenBerg, Kopacki  \&
  Ferguson}{Denissenkov et~al.}{2017}]{denissenkov_constraints_2017}
Denissenkov P.~A.,  VandenBerg D.~A.,  Kopacki G.,   Ferguson J.~W.,  2017,
  \mn@doi [\apj] {10.3847/1538-4357/aa92c9}, 849, 159

\bibitem[\protect\citeauthoryear{Di~Criscienzo, Marconi  \&
  Caputo}{Di~Criscienzo et~al.}{2004}]{di_criscienzo_rr_2004}
Di~Criscienzo M.,  Marconi M.,   Caputo F.,  2004, \mn@doi [\apj]
  {10.1086/422742}, 612, 1092

\bibitem[\protect\citeauthoryear{Di~Criscienzo et~al.,}{Di~Criscienzo
  et~al.}{2011}]{di_criscienzo_new_2011}
Di~Criscienzo M.,  et~al., 2011, \mn@doi [The Astronomical Journal]
  {10.1088/0004-6256/141/3/81}, 141, 81

\bibitem[\protect\citeauthoryear{Fabrizio et~al.,}{Fabrizio
  et~al.}{2019}]{fabrizio_use_2019}
Fabrizio M.,  et~al., 2019, \mn@doi [\apj] {10.3847/1538-4357/ab3977}, 882, 169

\bibitem[\protect\citeauthoryear{Fadeyev}{Fadeyev}{2019}]{fadeyev_period_2019}
Fadeyev Y.~A.,  2019, \mn@doi [Astronomy Letters] {10.1134/S1063773719060021},
  45, 353

\bibitem[\protect\citeauthoryear{Givens \& Pilachowski}{Givens \&
  Pilachowski}{2016}]{givens_abundance_2016}
Givens R.~A.,  Pilachowski C.~A.,  2016, \mn@doi [Publications of the
  Astronomical Society of the Pacific] {10.1088/1538-3873/128/970/124203}, 128,
  124203

\bibitem[\protect\citeauthoryear{Harris}{Harris}{1996}]{harris_catalog_1996}
Harris W.~E.,  1996, \mn@doi [\aj] {10.1086/118116}, 112, 1487

\bibitem[\protect\citeauthoryear{Harris}{Harris}{2010}]{harris_new_2010}
Harris W.~E.,  2010, \mn@doi [arXiv e-prints] {10.48550/arXiv.1012.3224}, p.
  arXiv:1012.3224

\bibitem[\protect\citeauthoryear{Hartman, Kaluzny, Szentgyorgyi  \&
  Stanek}{Hartman et~al.}{2005}]{hartman_bvi_2005}
Hartman J.~D.,  Kaluzny J.,  Szentgyorgyi A.,   Stanek K.~Z.,  2005, \mn@doi
  [\aj] {10.1086/427252}, 129, 1596

\bibitem[\protect\citeauthoryear{Hornik}{Hornik}{1991}]{hornik_approximation_1991}
Hornik K.,  1991, \mn@doi [Neural Networks]
  {https://doi.org/10.1016/0893-6080(91)90009-T}, 4, 251

\bibitem[\protect\citeauthoryear{Hornik, Stinchcombe  \& White}{Hornik
  et~al.}{1989}]{hornik_multilayer_1989}
Hornik K.,  Stinchcombe M.,   White H.,  1989, \mn@doi [Neural Networks]
  {https://doi.org/10.1016/0893-6080(89)90020-8}, 2, 359

\bibitem[\protect\citeauthoryear{Iben \& Huchra}{Iben \&
  Huchra}{1971}]{iben_comments_1971}
Iben I. J.,  Huchra J.,  1971, \aap, 14, 293

\bibitem[\protect\citeauthoryear{Inno et~al.,}{Inno
  et~al.}{2015}]{inno_new_2015}
Inno L.,  et~al., 2015, \mn@doi [\aap] {10.1051/0004-6361/201424396}, 576, A30

\bibitem[\protect\citeauthoryear{Johnson, Kraft, Pilachowski, Sneden, Ivans  \&
  Benman}{Johnson et~al.}{2005}]{johnson_235_2005}
Johnson C.~I.,  Kraft R.~P.,  Pilachowski C.~A.,  Sneden C.,  Ivans I.~I.,
  Benman G.,  2005, \mn@doi [Publications of the Astronomical Society of the
  Pacific] {10.1086/497435}, 117, 1308

\bibitem[\protect\citeauthoryear{Jones, Carney  \& Fulbright}{Jones
  et~al.}{1996}]{jones_template_1996}
Jones R.~V.,  Carney B.~W.,   Fulbright J.~P.,  1996, \mn@doi [\pasp]
  {10.1086/133809}, 108, 877

\bibitem[\protect\citeauthoryear{Jurcsik}{Jurcsik}{2019}]{jurcsik_blazhko-type_2019}
Jurcsik J.,  2019, \mn@doi [\mnras] {10.1093/mnras/stz2498}, 490, 80

\bibitem[\protect\citeauthoryear{Jurcsik \& Kovács}{Jurcsik \&
  Kovács}{1996}]{jurcsik_determination_1996}
Jurcsik J.,  Kovács G.,  1996, Astronomy and Astrophysics, 312, 111

\bibitem[\protect\citeauthoryear{Jurcsik et~al.,}{Jurcsik
  et~al.}{2012}]{jurcsik_long-term_2012}
Jurcsik J.,  et~al., 2012, \mn@doi [\mnras] {10.1111/j.1365-2966.2011.19868.x},
  419, 2173

\bibitem[\protect\citeauthoryear{Jurcsik et~al.,}{Jurcsik
  et~al.}{2015}]{jurcsik_overtone_2015}
Jurcsik J.,  et~al., 2015, \mn@doi [The Astrophysical Journal Supplement
  Series] {10.1088/0067-0049/219/2/25}, 219, 25

\bibitem[\protect\citeauthoryear{Jurcsik et~al.,}{Jurcsik
  et~al.}{2017}]{jurcsik_photometric_2017}
Jurcsik J.,  et~al., 2017, \mn@doi [\mnras] {10.1093/mnras/stx382}, 468, 1317

\bibitem[\protect\citeauthoryear{Jurcsik, Hajdu, Dékány, Nuspl, Catelan  \&
  Grebel}{Jurcsik et~al.}{2018}]{jurcsik_blazhko_2018}
Jurcsik J.,  Hajdu G.,  Dékány I.,  Nuspl J.,  Catelan M.,   Grebel E.~K.,
  2018, \mn@doi [\mnras] {10.1093/mnras/sty112}, 475, 4208

\bibitem[\protect\citeauthoryear{Kaluzny, Kubiak, Szymanski, Udalski,
  Krzeminski  \& Mateo}{Kaluzny et~al.}{1997}]{kaluzny_optical_1997}
Kaluzny J.,  Kubiak M.,  Szymanski M.,  Udalski A.,  Krzeminski W.,   Mateo M.,
   1997, \mn@doi [\aaps] {10.1051/aas:1997376}, 125, 343

\bibitem[\protect\citeauthoryear{Kingma \& Ba}{Kingma \&
  Ba}{2014}]{kingma_adam_2014}
Kingma D.~P.,  Ba J.,  2014, Adam: {A} {Method} for {Stochastic}
  {Optimization}, \mn@doi{10.48550/arXiv.1412.6980}, \url
  {https://ui.adsabs.harvard.edu/abs/2014arXiv1412.6980K}

\bibitem[\protect\citeauthoryear{Kovacs \& Kanbur}{Kovacs \&
  Kanbur}{1998}]{kovacs_modelling_1998}
Kovacs G.,  Kanbur S.~M.,  1998, \mn@doi [\mnras]
  {10.1046/j.1365-8711.1998.01271.x}, 295, 834

\bibitem[\protect\citeauthoryear{Kovacs \& Zsoldos}{Kovacs \&
  Zsoldos}{1995}]{kovacs_new_1995}
Kovacs G.,  Zsoldos E.,  1995, Astronomy and Astrophysics, 293, L57

\bibitem[\protect\citeauthoryear{Kovács \& Jurcsik}{Kovács \&
  Jurcsik}{1996}]{kovacs_light_1996}
Kovács G.,  Jurcsik J.,  1996, The Astrophysical Journal, 466, L17

\bibitem[\protect\citeauthoryear{Kumar, Bhardwaj, Singh, Das, Marconi, Kanbur
  \& Prugniel}{Kumar et~al.}{2023}]{kumar_predicting_2023}
Kumar N.,  Bhardwaj A.,  Singh H.~P.,  Das S.,  Marconi M.,  Kanbur S.~M.,
  Prugniel P.,  2023, \mn@doi [\mnras] {10.1093/mnras/stad937}, 522, 1504

\bibitem[\protect\citeauthoryear{Kunder et~al.,}{Kunder
  et~al.}{2013}]{kunder_rr_2013}
Kunder A.,  et~al., 2013, \mn@doi [The Astronomical Journal]
  {10.1088/0004-6256/146/5/119}, 146, 119

\bibitem[\protect\citeauthoryear{Kunder et~al.,}{Kunder
  et~al.}{2018}]{kunder_impact_2018}
Kunder A.,  et~al., 2018, \mn@doi [ßr] {10.1007/s11214-018-0519-0}, 214, 90

\bibitem[\protect\citeauthoryear{Landolt}{Landolt}{1992}]{landolt_ubvri_1992}
Landolt A.~U.,  1992, \mn@doi [\aj] {10.1086/116242}, 104, 340

\bibitem[\protect\citeauthoryear{Lee \& Sneden}{Lee \&
  Sneden}{2021}]{lee_multiple_2021}
Lee J.-W.,  Sneden C.,  2021, \mn@doi [\apj] {10.3847/1538-4357/abd948}, 909,
  167

\bibitem[\protect\citeauthoryear{Lenz \& Breger}{Lenz \&
  Breger}{2005}]{lenz_period04_2005}
Lenz P.,  Breger M.,  2005, \mn@doi [Communications in Asteroseismology]
  {10.1553/cia146s53}, 146, 53

\bibitem[\protect\citeauthoryear{Li, Qian  \& Zhu}{Li
  et~al.}{2018}]{li_period_2018}
Li L.-J.,  Qian S.-B.,   Zhu L.-Y.,  2018, \mn@doi [The Astrophysical Journal]
  {10.3847/1538-4357/aad32f}, 863, 151

\bibitem[\protect\citeauthoryear{Lomb}{Lomb}{1976}]{lomb_least-squares_1976}
Lomb N.~R.,  1976, Astrophysics and space science, 39, 447

\bibitem[\protect\citeauthoryear{Longmore, Fernley  \& Jameson}{Longmore
  et~al.}{1986}]{longmore_rr_1986}
Longmore A.~J.,  Fernley J.~A.,   Jameson R.~F.,  1986, \mn@doi [\mnras]
  {10.1093/mnras/220.2.279}, 220, 279

\bibitem[\protect\citeauthoryear{Longmore, Dixon, Skillen, Jameson  \&
  Fernley}{Longmore et~al.}{1990}]{longmore_globular_1990}
Longmore A.~J.,  Dixon R.,  Skillen I.,  Jameson R.~F.,   Fernley J.~A.,  1990,
  \mnras, 247, 684

\bibitem[\protect\citeauthoryear{Madore}{Madore}{1982}]{madore_Period-Luminosity_1982}
Madore B.~F.,  1982, \mn@doi [The Astrophysical Journal] {10.1086/159659}, 253,
  575

\bibitem[\protect\citeauthoryear{Marconi \& Degl'Innocenti}{Marconi \&
  Degl'Innocenti}{2007}]{marconi_modeling_2007}
Marconi M.,  Degl'Innocenti S.,  2007, \mn@doi [\aap]
  {10.1051/0004-6361:20065840}, 474, 557

\bibitem[\protect\citeauthoryear{Marconi, Caputo, Di~Criscienzo  \&
  Castellani}{Marconi et~al.}{2003}]{marconi_rr_2003}
Marconi M.,  Caputo F.,  Di~Criscienzo M.,   Castellani M.,  2003, \mn@doi [The
  Astrophysical Journal] {10.1086/377641}, 596, 299

\bibitem[\protect\citeauthoryear{Marconi et~al.,}{Marconi
  et~al.}{2015}]{marconi_new_2015}
Marconi M.,  et~al., 2015, \mn@doi [The Astrophysical Journal]
  {10.1088/0004-637X/808/1/50}, 808, 50

\bibitem[\protect\citeauthoryear{Molnár et~al.,}{Molnár
  et~al.}{2021}]{molnar_first_2021}
Molnár L.,  et~al., 2021, The Astrophysical Journal Supplement Series, 258, 8

\bibitem[\protect\citeauthoryear{Mullen et~al.,}{Mullen
  et~al.}{2023}]{mullen_rr_2023}
Mullen J.~P.,  et~al., 2023, \mn@doi [The Astrophysical Journal]
  {10.3847/1538-4357/acb20a}, 945, 83

\bibitem[\protect\citeauthoryear{Muraveva et~al.,}{Muraveva
  et~al.}{2015}]{muraveva_new_2015}
Muraveva T.,  et~al., 2015, \mn@doi [The Astrophysical Journal]
  {10.1088/0004-637X/807/2/127}, 807, 127

\bibitem[\protect\citeauthoryear{Neeley et~al.,}{Neeley
  et~al.}{2019}]{neeley_standard_2019}
Neeley J.~R.,  et~al., 2019, \mn@doi [\mnras] {10.1093/mnras/stz2814}, 490,
  4254

\bibitem[\protect\citeauthoryear{Nemec et~al.,}{Nemec
  et~al.}{2011}]{nemec_fourier_2011}
Nemec J.~M.,  et~al., 2011, \mn@doi [\mnras]
  {10.1111/j.1365-2966.2011.19317.x}, 417, 1022

\bibitem[\protect\citeauthoryear{O'Malley, Bursztein, Long, Chollet, Jin,
  Invernizzi  \& {others}}{O'Malley et~al.}{2019}]{omalley_kerastuner_2019}
O'Malley T.,  Bursztein E.,  Long J.,  Chollet F.,  Jin H.,  Invernizzi L.,
  {others} 2019, {KerasTuner}, \url {https://github.com/keras-team/keras-tuner}

\bibitem[\protect\citeauthoryear{Oosterhoff}{Oosterhoff}{1939}]{oosterhoff_remarks_1939}
Oosterhoff P.~T.,  1939, The Observatory, 62, 104

\bibitem[\protect\citeauthoryear{Paxton, Bildsten, Dotter, Herwig, Lesaffre  \&
  Timmes}{Paxton et~al.}{2011}]{paxton_modules_2011}
Paxton B.,  Bildsten L.,  Dotter A.,  Herwig F.,  Lesaffre P.,   Timmes F.,
  2011, \mn@doi [The Astrophysical Journal Supplement Series]
  {10.1088/0067-0049/192/1/3}, 192, 3

\bibitem[\protect\citeauthoryear{Paxton et~al.,}{Paxton
  et~al.}{2013}]{paxton_modules_2013}
Paxton B.,  et~al., 2013, \mn@doi [The Astrophysical Journal Supplement Series]
  {10.1088/0067-0049/208/1/4}, 208, 4

\bibitem[\protect\citeauthoryear{Paxton et~al.,}{Paxton
  et~al.}{2015}]{paxton_modules_2015}
Paxton B.,  et~al., 2015, \mn@doi [The Astrophysical Journal Supplement Series]
  {10.1088/0067-0049/220/1/15}, 220, 15

\bibitem[\protect\citeauthoryear{Paxton et~al.,}{Paxton
  et~al.}{2018}]{paxton_modules_2018}
Paxton B.,  et~al., 2018, \mn@doi [The Astrophysical Journal Supplement Series]
  {10.3847/1538-4365/aaa5a8}, 234, 34

\bibitem[\protect\citeauthoryear{Paxton et~al.,}{Paxton
  et~al.}{2019}]{paxton_modules_2019}
Paxton B.,  et~al., 2019, \mn@doi [The Astrophysical Journal Supplement Series]
  {10.3847/1538-4365/ab2241}, 243, 10

\bibitem[\protect\citeauthoryear{Pritzl, Smith, Catelan  \& Sweigart}{Pritzl
  et~al.}{2001}]{pritzl_variable_2001}
Pritzl B.~J.,  Smith H.~A.,  Catelan M.,   Sweigart A.~V.,  2001, \mn@doi [\aj]
  {10.1086/323447}, 122, 2600

\bibitem[\protect\citeauthoryear{Pritzl, Smith, Catelan  \& Sweigart}{Pritzl
  et~al.}{2002}]{pritzl_variable_2002}
Pritzl B.~J.,  Smith H.~A.,  Catelan M.,   Sweigart A.~V.,  2002, \mn@doi [\aj]
  {10.1086/341381}, 124, 949

\bibitem[\protect\citeauthoryear{Rood}{Rood}{1973}]{rood_metal-poor_1973}
Rood R.~T.,  1973, \mn@doi [\apj] {10.1086/152373}, 184, 815

\bibitem[\protect\citeauthoryear{{Savino, A.}, {Koch, A.}, {Prudil, Z.},
  {Kunder, A.}  \& {Smolec, R.}}{{Savino, A.} et~al.}{2020}]{savino_a_age_2020}
{Savino, A.} {Koch, A.} {Prudil, Z.} {Kunder, A.}  {Smolec, R.} 2020, \mn@doi
  [A\&A] {10.1051/0004-6361/202038305}, 641, A96

\bibitem[\protect\citeauthoryear{Scargle}{Scargle}{1982}]{scargle_studies_1982}
Scargle J.~D.,  1982, \mn@doi [The Astrophysical Journal] {10.1086/160554},
  263, 835

\bibitem[\protect\citeauthoryear{Schlegel, Finkbeiner  \& Davis}{Schlegel
  et~al.}{1998}]{schlegel_maps_1998}
Schlegel D.~J.,  Finkbeiner D.~P.,   Davis M.,  1998, \mn@doi [The
  Astrophysical Journal] {10.1086/305772}, 500, 525

\bibitem[\protect\citeauthoryear{Sesar et~al.,}{Sesar
  et~al.}{2009}]{sesar_light_2009}
Sesar B.,  et~al., 2009, The Astrophysical Journal, 708, 717

\bibitem[\protect\citeauthoryear{Shapley}{Shapley}{1916}]{shapley_changes_1916}
Shapley H.,  1916, \mn@doi [{\textbackslash}apj] {10.1086/142246}, 43, 217

\bibitem[\protect\citeauthoryear{Siegel, Porterfield, Balzer  \& Hagen}{Siegel
  et~al.}{2015}]{siegel_swift_2015}
Siegel M.~H.,  Porterfield B.~L.,  Balzer B.~G.,   Hagen L. M.~Z.,  2015,
  \mn@doi [\aj] {10.1088/0004-6256/150/4/129}, 150, 129

\bibitem[\protect\citeauthoryear{Simon \& Clement}{Simon \&
  Clement}{1993}]{simon_provisional_1993}
Simon N.~R.,  Clement C.~M.,  1993, \mn@doi [The Astrophysical Journal]
  {10.1086/172771}, 410, 526

\bibitem[\protect\citeauthoryear{Smolec \& Moskalik}{Smolec \&
  Moskalik}{2008}]{smolec_convective_2008}
Smolec R.,  Moskalik P.,  2008, \mn@doi [Acta Astronomica]
  {10.48550/arXiv.0809.1979}, 58, 193

\bibitem[\protect\citeauthoryear{Sneden, Kraft, Guhathakurta, Peterson  \&
  Fulbright}{Sneden et~al.}{2004}]{sneden_chemical_2004}
Sneden C.,  Kraft R.~P.,  Guhathakurta P.,  Peterson R.~C.,   Fulbright J.~P.,
  2004, \mn@doi [\aj] {10.1086/381907}, 127, 2162

\bibitem[\protect\citeauthoryear{Sollima, Cacciari  \& Valenti}{Sollima
  et~al.}{2006}]{sollima_rr_2006}
Sollima A.,  Cacciari C.,   Valenti E.,  2006, \mn@doi [\mnras]
  {10.1111/j.1365-2966.2006.10962.x}, 372, 1675

\bibitem[\protect\citeauthoryear{Soszyński, Gieren  \&
  Pietrzyński}{Soszyński et~al.}{2005}]{soszynski_mean_2005}
Soszyński I.,  Gieren W.,   Pietrzyński G.,  2005, Publications of the
  Astronomical Society of the Pacific, 117, 823

\bibitem[\protect\citeauthoryear{Stetson}{Stetson}{1987}]{stetson_daophot_1987}
Stetson P.~B.,  1987, \mn@doi [\pasp] {10.1086/131977}, 99, 191

\bibitem[\protect\citeauthoryear{Stetson}{Stetson}{1994}]{stetson_center_1994}
Stetson P.~B.,  1994, \mn@doi [\pasp] {10.1086/133378}, 106, 250

\bibitem[\protect\citeauthoryear{Stetson et~al.,}{Stetson
  et~al.}{2014a}]{stetson_optical_2014}
Stetson P.~B.,  et~al., 2014a, \mn@doi [\pasp] {10.1086/677195}, 126, 521

\bibitem[\protect\citeauthoryear{Stetson, Fiorentino, Bono, Bernard, Monelli,
  Iannicola, Gallart  \& Ferraro}{Stetson
  et~al.}{2014b}]{stetson_homogeneous_2014}
Stetson P.~B.,  Fiorentino G.,  Bono G.,  Bernard E.~J.,  Monelli M.,
  Iannicola G.,  Gallart C.,   Ferraro I.,  2014b, \mn@doi [\pasp]
  {10.1086/677352}, 126, 616

\bibitem[\protect\citeauthoryear{Stetson, Pancino, Zocchi, Sanna  \&
  Monelli}{Stetson et~al.}{2019}]{stetson_homogeneous_2019}
Stetson P.~B.,  Pancino E.,  Zocchi A.,  Sanna N.,   Monelli M.,  2019, \mn@doi
  [Monthly Notices of the Royal Astronomical Society] {10.1093/mnras/stz585},
  485, 3042

\bibitem[\protect\citeauthoryear{Szeidl, Hurta, Jurcsik, Clement  \&
  Lovas}{Szeidl et~al.}{2011}]{szeidl_long-term_2011}
Szeidl B.,  Hurta Z.,  Jurcsik J.,  Clement C.,   Lovas M.,  2011, \mn@doi
  [Monthly Notices of the Royal Astronomical Society]
  {10.1111/j.1365-2966.2010.17815.x}, 411, 1744

\bibitem[\protect\citeauthoryear{{Valcarce, A. A. R.} \& {Catelan,
  M.}}{{Valcarce, A. A. R.} \& {Catelan,
  M.}}{2008}]{valcarce_a_a_r_semi-empirical_2008}
{Valcarce, A. A. R.} {Catelan, M.} 2008, \mn@doi [A\&A]
  {10.1051/0004-6361:20078231}, 487, 185

\bibitem[\protect\citeauthoryear{VandenBerg, Denissenkov  \&
  Catelan}{VandenBerg et~al.}{2016}]{vandenberg_constraints_2016}
VandenBerg D.~A.,  Denissenkov P.~A.,   Catelan M.,  2016, \mn@doi [\apj]
  {10.3847/0004-637X/827/1/2}, 827, 2

\bibitem[\protect\citeauthoryear{VanderPlas \& Ivezic}{VanderPlas \&
  Ivezic}{2015}]{vanderplas_periodograms_2015}
VanderPlas J.~T.,  Ivezic v.,  2015, The Astrophysical Journal, 812, 18

\bibitem[\protect\citeauthoryear{Wallerstein, Kovtyukh  \&
  Andrievsky}{Wallerstein et~al.}{2009}]{wallerstein_carbon-rich_2009}
Wallerstein G.,  Kovtyukh V.~V.,   Andrievsky S.~M.,  2009, \mn@doi [\apjl]
  {10.1088/0004-637X/692/2/L127}, 692, L127

\makeatother
\end{thebibliography}




\appendix 





\section{Training of ANN}\label{sec:apndx_train_ANN}
\begin{figure}
    \centering
    \includegraphics[scale=0.95]{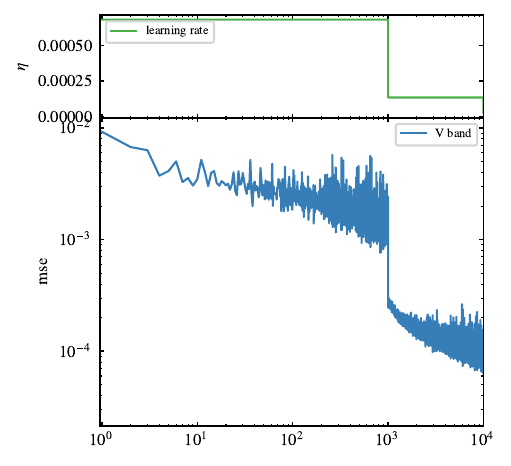}
    \caption{The training curves for $V$ band \emph{RRab-Net}.}
    \label{fig:train_curve}
\end{figure}

We used the architecture outlined in Table \ref{tab:network_architecture} to build the \emph{RRab-Net} for the $V$ band. This architecture selection followed a Grid search for hyperparameters, as explained in Section \ref{sec:physical_parameters}. It is essential to note that high learning rates can induce oscillations in the learning curve, which can be reduced by lowering the learning parameter. However, starting training with a low $\eta$ can negatively impact network performance. To address this, we used a `piece-wise decay' approach, where the current learning rate is divided by a constant number (`$\delta$') when the epoch crosses successive powers of $10$. After experimenting with different values, we chose $\delta = 5$. This ensures that the loss decreases gradually as training progresses. We trained the network for $10,000$ epochs and achieved a minimum MSE of $3.02 \times 10^{-5}$. The final learning curve and learning parameter ($\eta$) are shown in Fig.~\ref{fig:train_curve}.

\begin{table}
    \centering
    \caption{The adopted network architecture of the \emph{RRab-Net} after hyperparameter optimisation using a random search algorithm.}
    \resizebox{\linewidth}{!}{%
    \begin{tabular}{cll}
        \hline 
        S. N. & Name of hyperparameter & Value  \\
        \hline 
        1 & Number of hidden layers &  3 \\
        2 & No  of neurons in hidden layers & [128, 64, 64]   \\
        3 & Optimizer & `\texttt{adam}' \citep{kingma_adam_2014} \\
        4 & Learning rate  & $6.7984\times10^{-4}$ (decreasing \\
          &  & with number of epochs)  \\
        5 & Activation function & `\texttt{tanh}' \\
        6 & Weights initialisation & `\texttt{GlorotUniform}' \\
        \hline
    \end{tabular}
    }
    \label{tab:network_architecture}
\end{table}

\section{Additional Template Fit Light curves}
\begin{figure*}
    \centering
    \includegraphics{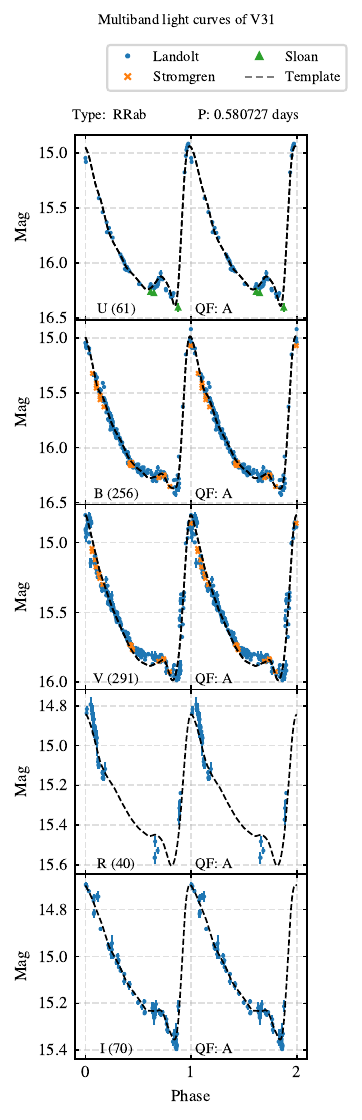}
    \includegraphics{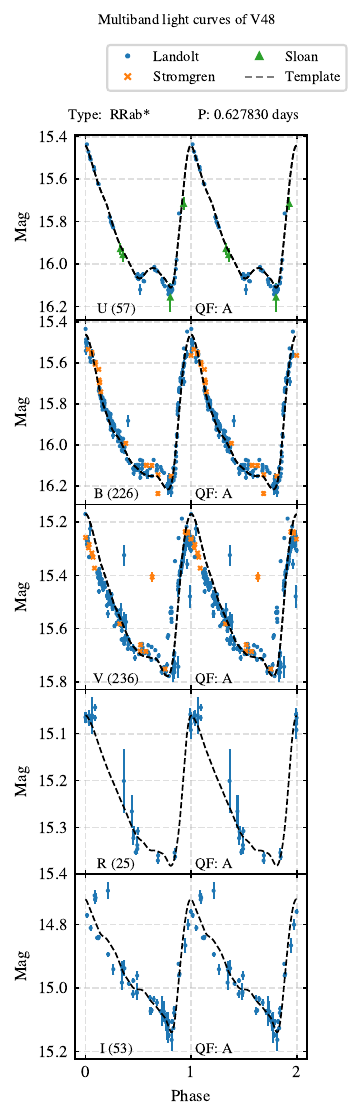}
    \caption{The left panel presents the $U$, $B$, $V$, $R$, and $I$ band light curves of the V31 (non-Blazhko RRab) star. The period of this star is 0.580727 days. The folded light curves of the Blazhko RRab star V48 are displayed in the right panel, with a primary period of 0.627830 days. The adopted nomenclature for this figure is same as Fig. \ref{fig:template_fit_rrab}.}
    \label{fig:apndx_tmplt_fit_rrab}
\end{figure*}

\begin{figure*}
    \centering
    \includegraphics{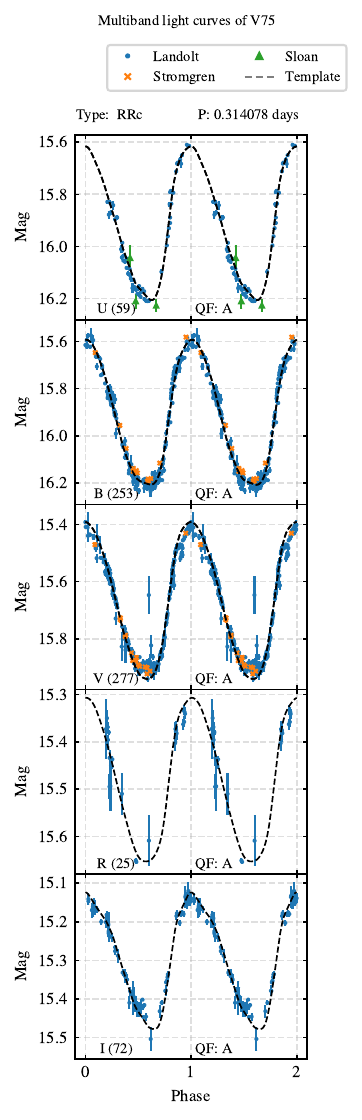}
    \includegraphics{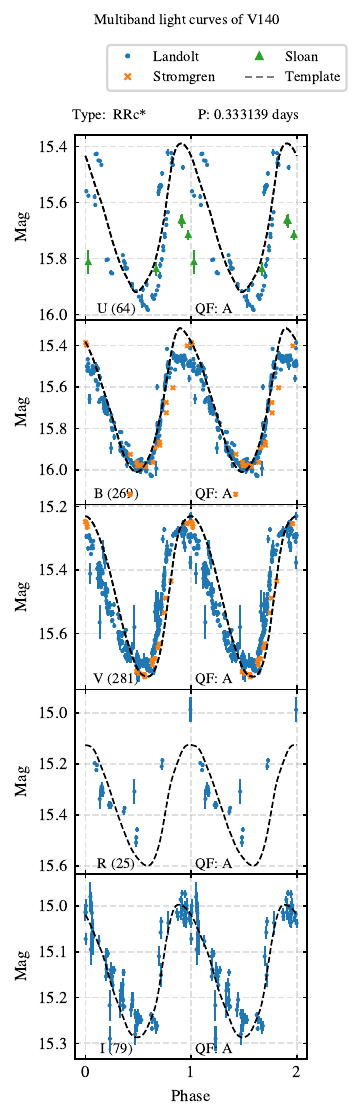}
    \caption{The left panel presents the $U$, $B$, $V$, $R$, and $I$ band light curves of the V75 (non-Blazhko RRc) star. The period of this star is 0.314078 days. The folded light curves of the Blazhko RRc star V140 are displayed in the right panel, with a primary period of 0.333139 days. The adopted nomenclature for this figure is same as Fig. \ref{fig:template_fit_rrab}.} 
    \label{fig:apndx_tmplt_fit_rrc}
\end{figure*}

We present two additional figures showcasing the light curves of RR Lyrae stars in different filter bands. Fig. \ref{fig:apndx_tmplt_fit_rrab} displays the light curves of the non-Blazhko RRab star V31 (left panel) and the Blazhko RRab star V48 (right panel). The period of V31 is 0.580727 days, while V48 exhibits a primary period of 0.627830 days. Similarly, Fig. \ref{fig:apndx_tmplt_fit_rrc} illustrates the light curves of the non-Blazhko RRc star V75 (left panel) and the Blazhko RRc star V140 (right panel). The period of V75 is 0.314078 days, whereas V140 demonstrates a primary period of 0.333139 days. Both figures follow the same nomenclature as depicted in Fig. \ref{fig:apndx_tmplt_fit_rrab} for clarity and consistency. 

\bsp	
\label{lastpage}


\end{document}